\documentclass{emulateapj}

\def\arcdeg{\hbox{$^\circ$}}
\def\arcmin{\hbox{$^\prime$}}
\def\arcsec{\hbox{$^{\prime\prime}$}}
\def\deg2{\hbox{$\rm deg^{2}$}}

\lefthead{Drake et~al.}
\righthead{Catalina Real-time Transient Survey}

\begin{document}
\title{First Results from the Catalina Real-time Transient Survey}


\author{A.J. Drake\altaffilmark{1}, S.~G. Djorgovski\altaffilmark{1}, A. Mahabal\altaffilmark{1}, E. Beshore\altaffilmark{2},
S. Larson\altaffilmark{2}, M.~J. Graham\altaffilmark{1},\\ R. Williams\altaffilmark{1}, E. Christensen\altaffilmark{3},
M. Catelan\altaffilmark{4,5,6,7}, A. Boattini\altaffilmark{2}, A. Gibbs\altaffilmark{2}, R. Hill\altaffilmark{2}\\ 
and R. Kowalski\altaffilmark{2}
}

\altaffiltext{1}{California Institute of Technology, 1200 E. California Blvd, CA 91225, USA}
\altaffiltext{2}{The University of Arizona, Dept. of Planetary Sciences,  Lunar and Planetary Laboratory, 1629 E. University Blvd, Tucson AZ 85721, USA}
\altaffiltext{3}{Gemini Observatory, Casilla 603, La Serena, CL, Chile}
\altaffiltext{4}{John Simon Guggenheim Memorial Foundation Fellow}
\altaffiltext{5}{Pontificia Universidad Cat\'olica de Chile, Departamento de Astronom\'ia y Astrof\'isica, 
Av. Vicu\~na Mackena 4860, 782-0436 Macul, Santiago, Chile}
\altaffiltext{6}{NASA Goddard Space Flight Center, Exploration of the Universe Division,
Code 667, Greenbelt, MD 20771}
\altaffiltext{7}{Catholic University of America, Department of Physics, 200 Hannan Hall,Washington, DC 20064}


\date{Received 2008 / Accepted 2008}


\begin{abstract}
  
  We report on the results from the first six months of the Catalina Real-time Transient Survey (CRTS). 
  In order to search for optical transients with timescales of minutes to years, the CRTS analyses data 
  from the Catalina Sky Survey which repeatedly covers twenty six thousand of square degrees on the sky. 
  The CRTS provides a public stream of transients that are bright enough to be followed
  up using small telescopes.  Since the beginning of the survey, all CRTS transients have been made
  available to astronomers around the world in real-time using HTML tables, RSS feeds and VOEvents.  As part of our 
  public outreach program the detections are now also available in KML through Google Sky.
  
  The initial discoveries include over 350 unique optical transients rising more than two magnitudes from past
  measurements.  Sixty two of these are classified as supernovae, based on light curves, prior deep imaging and
  spectroscopic data.  Seventy seven are due to cataclysmic variables (only 13 previously known), while an additional
  100 transients were too infrequently sampled to distinguish between faint CVs and SNe.  The remaining optical
  transients include AGN, Blazars, high proper motions stars, highly variable stars (such as UV Ceti stars) and
  transients of an unknown nature.  Our results suggest that there is a large population of SNe missed by many current
  supernova surveys because of selection biases. These objects appear to be associated with faint host galaxies.  We
  also discuss the unexpected discovery of white dwarf binary systems through dramatic eclipses.

\end{abstract}

\section{Introduction}

Time-domain astronomy is one of the most rapidly emerging areas of astronomy (Paczy\'nski 2000). Several large 
experiments (LSST, Ivezic et~al. 2008; PanSTARRS, Hodapp et~al. 2004; 
SkyMapper, Keller et~al. 2007) are set to make a major impact in the near future
by covering thousands of square degrees with targets ranging from distant supernovae to near earth asteroids.
Past time-domain surveys have mainly concentrated on areas of tens to hundreds of square degrees and have typically
searched for specific types of astronomical transients such as microlensing (MACHO, Alcock et~al. 1993; OGLE,
Udalski et~al. 1994; EROS, Aubourg et~al. 1995), GRB afterglows (ROTSE; Akerloff et~al. 2000) and supernovae 
(KAIT; Filippenko et~al. 2001). 

Deep surveys for variability have been carried out over small areas ($< 25$ $\rm deg^2$)
by the Subaru/XMM-Newton Deep Survey (SXDS; Morokuma et al. 2008),
the Deep Lensing Survey (DLS; Becker et al. 2004) and the 
Faint Sky Variability Survey (FSVS; Huber et al. 2006) wheras large-area surveys, such as the Sloan Digital Sky 
Survey (SDSS; York et~al. 2000), the Two Micron All Sky Survey (2MASS; Skrutskie et~al. 2006), and the Galaxy 
Evolution Explorer (GALEX; Martin et~al. 2005) have generally not been synoptic.
One survey intermediate between past deep surveys, and wide future surveys was carried out by 
the SDSS consortium to discover type Ia supernovae. This survey covered 300 $\rm deg^2$ in 5 optical bands 
to $r \sim 23$ during nine months spaced over 2005 to 2007 (Sesar et~al. 2007).

\subsection{Current Transient Surveys}

Current wide-field transient surveys include the Robotic Optical Transient Search Experiment (ROTSE-III; Akerlof et
al. 2003) and the All Sky Automated Survey (ASAS-3; Pojmanski 2001). However, like many other surveys, both are targeted to the 
discovery and characterization of GRB afterglows.
ASAS-3 surveys $\sim$ 30,000 square degrees of the southern sky to  $V\sim13.5$, every 2 nights (Pojmanski 2001),
while the four telescopes of the ROTSE-III survey routinely cover 1260 square degrees to $R\sim17.5$ 
(Rykoff et al. 2005; Yost et al. 2007).

In the time between targeted surveys, and future wide, deep, transient surveys, two experiments have been
searching tens of thousands of square degrees of the sky for transients; the Palomar-Quest digital synoptic 
sky survey (PQ\footnote{http://palquest.org}, Djorgovski et al. 2008a), and the Catalina Real-time 
Transient Survey (CRTS).  The PQ survey started analyzing driftscan data for optical transients in real-time in 
August 2006, while the CRTS began in November 2007. The transients discovered in PQ data will be presented 
in Mahabal et~al. (2008a).

Both CRTS and PQ experiments use a purpose-built pipeline for real-time transient detection, analysis and distribution
(Drake et~al. 2008a).  These surveys are now set to provide estimates of the rates and types of optical transients that
can be expected from future synoptic surveys.

In these experiments, unlike most current and past optical transient surveys, the results are 
made public within minutes of discovery using the IVOA standard VOEvent protocol and the VOEventNet
astronomical event distribution network (Drake et al. 2007a). In addition, VOEventNet, in collaboration with Google, 
makes transient astronomy available to the world in real-time via a dedicated layer in Google 
Sky\footnote{http://earth.google.com/sky}. Current events available through VOEventNet also include
MOA and OGLE microlensing, GCN gamma ray burst alerts, and PQ transients.

The CRTS routinely searches for Optical Transients (OTs) in data from the Catalina Sky Survey Schmidt telescope (CSS).
In this paper we present results from the first six months.  Firstly, we will discuss the observations and data analysis
($\S 2$). We will then list the current results ($\S 3$) and discuss some of the types of transients discovered.
Finally, we will summarize the current findings ($\S 4$).

\section{Observations and Data Reduction}

The Catalina Sky Survey\footnote{http://www.lpl.arizona.edu/css/} (CSS; Larson et~al. 2003) uses the 0.7m f/1.9 Catalina
Schmidt Telescope north of Tucson, Arizona to discover Near Earth Objects (NEOs) and Potentially Hazardous Asteroids
(PHAs).  This survey uses a single unfiltered 4k x 4k CCD with $2.5\arcsec$ pixels, giving an eight square degree field-of-view.
To date, CSS has found hundreds of PHAs, comets and other solar system objects, and currently leads the rate of NEA
discoveries\footnote{http://neo.jpl.nasa.gov/stats/}.  On a clear night the Catalina Schmidt typically
covers $\sim 1200$ squares degrees of sky in a sequence of four 30 second exposures. Currently observed fields cover 
$\sim$ 26,000 square degrees in the declination range $-30\arcdeg < \delta < 70\arcdeg$.  Observations are generally 
limited to Galactic latitudes $|b| > 10\arcdeg$ to prevent confusion caused by crowding. Weather permitting,
observations are made on 21 to 24 nights per lunation and typically reach V magnitudes from 19 to 20.
The sets of four CSS images are taken in sequence and evenly spaced over $\sim 30$ minutes.  This allows us to detect
transients varying on timescales from minutes to years.  In addition, the four image sequence provides a significant veto for
asteroids and artifacts that often cannot be distinguished from genuine rapid transients in pairs of exposures.

In the near future we hope to include two additional dedicated CSS telescopes in OT searches.
The 1.5m Mt.~Lemmon reflector near Tucson, with a one square degree field-of-view,
and the 0.5m Uppsala Schmidt at Siding Spring, Australia, with a 4.2 square degree field-of-view. 
All three telescopes operate in the same mode with similar cameras and reduction software. 

\subsection{Transient Detection}

The main goal of the CRTS thus far has been to discover and characterize variable source populations that exhibit
high-amplitude variablity on the timescales probed by CSS data.  In this paper we define optical transients as
objects that vary in brightness by greater than two magnitudes between past catalogs derived from high S/N co-added
images.  The OT characterization process undertaken here includes understanding known types of variable objects, as well
as searching for new kinds of OTs.  At the same time, the results of this survey haave been designed to deliver a transient
discovery stream that is available to the entire astronomical community as a testbed for VOEventNet technology.

The CSS uses standard SExtractor photometry software to produce object catalogs for each image as it is read out.  To
find transients we compare new detections against deep, clean source catalogs, rather than simply comparing these to
catalogs from earlier observations. This is necessary as image artifacts and differences in image depth lead to an
overwhelming number of unmatched detections.  The CSS detections are primarily matched with objects detected in past CSS
co-added data. Each co-add consists of the median combination of at least 20 images. These images reach sources
approximately two magnitudes fainter than the deepest individual CSS images (mag $\sim 22$).  Thus, very few objects are
missed in the reference catalog because of image artifacts or changes in depth and transparency. However, some objects
are missing in the catalogs because they are blended in the co-added images. For this reason the OT candidates are also
matched against objects in the higher resolution USNO-B, SDSS
and PQ surveys. 

An alternate method to catalog searches for finding transients is image subtraction.  The image subtraction technique
involves matching new observations to a high signal-to-noise template image before subtracting one from the other 
(Tomaney and Crotts 1996, Drake et~al. 1999). The resulting difference image is then searched for transients.  Although this increases the noise
associated with detections, this is a very effective way of revealing transient objects in the presence of significant
flux from blended constant sources.  For this reason subtraction has been particularly successful for dense stellar
fields observed in supernova searches such used by the SNfactory (Aldering et~al. 2002). However, this technique does
not work well at blue wavelengths in broad filters because of atmospheric differential refraction (Drake et~al. 1999).
The resulting difference images have bipolar residuals at the locations of the stars and often lead to false detections
and missed transients. The image subtraction process is also prone to fail, or produce artifacts, when images are taken
in bad observing conditions.  As catalog-based searches simply match lists of sources, they do not introduce new
artifacts into the images, making them more suitable for unsupervised real-time transient searches.

In the PQ survey we found that the detection of OTs in a single scan was hampered by the presence of large numbers of previously
undiscovered asteroids (Mahabal et~al. 2008a). By design, the longer temporal separation of CSS images means that almost
all asteroids can be detected from their motion between images. In addition, the requirement of positional coincidence
between images aids the removal of most image artifacts. The four image sequence used by CSS to discover NEOs also allow
us to search for objects that vary on timescales of minutes.

The structure of the CRTS data processing pipeline largely follows the PQ pipeline (Drake et~al. 2008a).  In short, all
the CSS catalogs are processed as the images are taken, on site.  Candidate OTs are filtered to remove detections
associated with artifacts (such as saturation spikes), and moving objects (such as asteroids and satellite trails).
After filtering, the transient cutout images, lightcurves and associated meta-data are posted to VOEventNet
where the OTS are posted on webpages and an RSS feeds\footnote{http://voeventnet.caltech.edu/feeds/Catalina.shtml}.
Additional image cutouts (from SDSS, PQ, and the DSS) are added to the information about each OT candidate by VOEventNet.
As one of the goals of CRTS is to enable rapid automated follow-up of short timescale transients by robotic telescopes,
VOEvent alerts are sent from the observatory as soon as the data is processed (approximately 5 minutes after the fourth 
observation in a sequence is obtained). 

After the filtering process $\sim 1$ source in 200,000 is detected as a significant OT. All of these candidates are sent 
as VOEvents. Currently $\sim 50\%$ of these are associated with genuine OTs. The remaining handful of transient 
candidates are screened by eye within minutes to hours of discovery and a updated list of transients is
posted\footnote{http://nesssi.cacr.caltech.edu/catalina/Allns.html}.

\section{Results}

In the first six months of the CRTS survey, images were searched covering a total $\sim$450,000 square degrees 
from $\sim 2000$, eight square degree fields. Over 350 OTs passed selection by eye. These included dozens of newly 
discovered cataclysmic variables (Drake et~al., 2007b, 2008b, 2008d, 2008f, 2008h; Djorgovski et~al. 2008b, 2008d, Glikman et~al. 2008),
flaring events (Drake et~al., 2008e; Djorgovski et~al. 2007),
nearby high-proper motion stars, Blazars (Mahabal et~al. 2008b), asteroids and comets. 
In addition, although CRTS has not focused on SNe, many OTs have been confirmed as supernovae, including 
SN 2007sr (Drake et~al. 2007b), a bright supernova in the Antennae galaxies.  Fainter SNe have also been discovered 
(SN 2008au, 2008av, 2008al, 2008ba, 2008bb, 2008bm, 2008cg, 2008ck, 2008dc, 2008dd, 2008de, 2008df, 2008dk (Drake
et~al. 2007a, 2007b, 2008c, 2008e, 2008f, 2008g, 2008h, 2008i; and Mahabal et~al. 2008b)).
Nine known SNe were also independently rediscovered by the CRTS pipeline (SNe 2006tf, 2007nm, 2007no, 
2007pu, 2007qv, 2008aq, 2008aw, 2008ax, 2008dk). 

Many types of optical transients are readily distinguished in a small number of images. For instance, the presence of
motion between images is the certain sign a transient is a nearby object. However, there are a number of cases where
classification based on a single set of images is difficult.
For instance, dwarf novae (DNe) type CVs can produce outbursts as large as 
eight magnitudes, rising to an absolute V-band magnitude of two (Harrison et al. 2004).
The presence of such large outbursts means that faint CV systems may be undetectable 
at quiescence. In a similar way supernova can rise more than 20 magnitudes and range
between $-20 < M_{V} < -16$ at peak brightness.
Supernovae are a common end point of stellar evolution and can occur in
undetected dwarf galaxies many magnitudes fainter than the SN itself. 
Although the absolute magnitudes of DNe and SNe are markedly different, in sparsely sampled 
data, both DNe and SNe associated with faint sources can be difficult to distinguish at discovery. 
For this reason transients associated with bright galaxies are more likely to be followed as SNe candidates. 
Clearly this kind of selection bias means that many SNe are never followed and this may also have led to errors in 
determinations of supernova rates. For instance, core collapse SNe (II, Ib/c) and type Ia SN are well known to occur 
at different rates depending on galaxy morphology (Mannucci et al. 2008). The fact that few SNe are followed in faint galaxies 
may mean that the types of SNe occurring in these settings is underestimated. 

DNe can readily be distinguished from SNe with spectroscopic observations.  However, the number of telescopes capable of
photometrically observing faint sources far outnumbers those which can spectroscopically monitor such sources.  When
spectroscopic observations are unavailable, other differences can be investigated.  For instance, DNe outbursts typically
last from few days to a couple of weeks, while SNe are bright for months. DNe outbursts take 1 to 3 days to rise to peak
brightness while most SNe take 10 to 20 days. DNe recur on scales of days to years while SNe are one-time events.  Both
the DNe system and their outbursts are usually very blue while SNe pass through a range of colours as they age (depending
on their type and host galaxy extinction).  DNe can also be distinguished from SNe using accurate measurement of source
colour over time (Poznanski et al. 2002).  Most CV systems can also be distinguished from variable stars on the giant
branch and main sequence stars in multi-band photometry. For instance, from blue $(g-r)$ and $(r-i)$ colours in SDSS data
(Kriscinunas, Margon, \& Szkody 1998).

\begin{figure}[ht]{
\epsscale{0.8}
\plotone{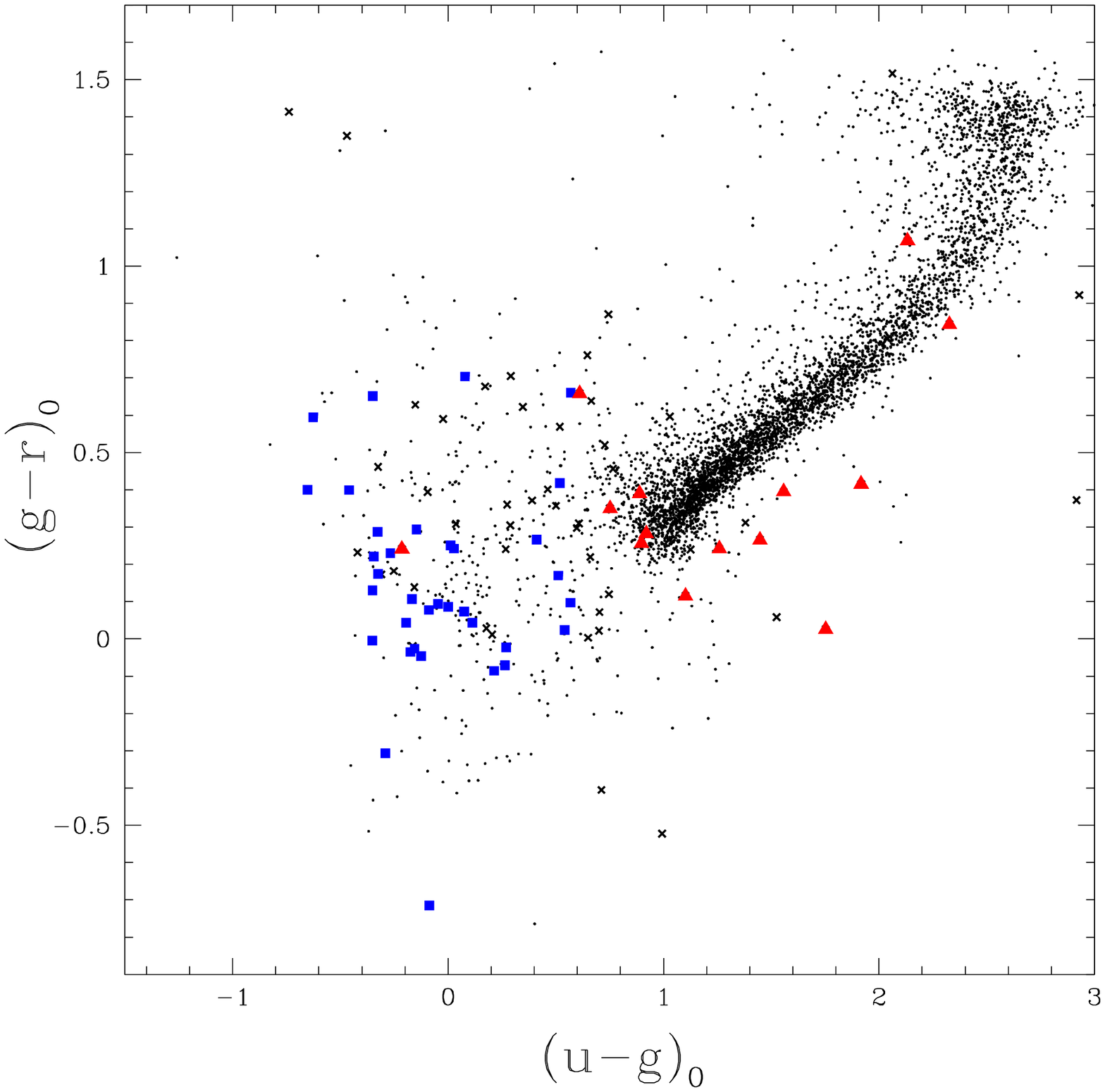}
\caption{\label{sncolour}
A color-color diagram of CRTS transients matching SDSS-DR6 sources.
Triangles: likely and confirmed SN. Squares: confirmed and likely dwarf novae.
Crosses: all other transients including possible SNe and DNe.
Dots show the de-reddened stellar locus.
}
}
\end{figure}

\placefigure{sncolour}

In Figure \ref{sncolour}, we present the de-reddened colour-colour diagram of all the CRTS transients that match SDSS
sources within $2.5\arcsec$. In this figure we have separated SNe and DNe candidates with different symbols
so that differences may be clearly seen. 
This figure shows that CVs are generally bluer than the SNe hosts in ($u-g$) colours. The ($u-g$)
colours are consistent with those expected for white dwarfs (Eisenstein et al.~2006) and CVs discovered in SDSS data
(Szkody et al.~2007).  Since this figure only includes matches within $2.5\arcsec$, it does not include the SNe
discovered which were associated with large nearby galaxies.  For these objects the offsets from the galactic centers
are much larger. The colour separation of the two populations seen in Figure \ref{sncolour} tells us that if we see
a source with $(u - g) < 0.7$ it is much more likely to be due to a CV than an SNe.
Indeed of the 213 new and known CVs measured by Szkody et al. (2002; 2003; 2004; 2005; 2006; 2007) in SDSS data, 
only ten had $(u - g) > 0.7$.

\subsection{Supernovae}

Supernovae are explosive end points of stellar evolution. In some cases these objects can be used as cosmological
standard candles that probe the depths of the expanding Universe and aid our understanding Dark Energy. Many groups 
are currently searching for nearby SNe (Lick Observatory Supernova Search - LOSS; Filipenko et al. 2001, The Nearby
Supernova Factory - SNfactory; Aldering et al. 2002, Chilean Automatic Supernova Search - CHASE; Pignata et al. 2007,
Monard 2003, Puckett, Langoussis, \& Marcus 2003).
Most of these groups only search for low-z SNe clearly associated with large, bright, galaxies
(Prieto et al. 2008)\footnote{http://www.supernovae.net}.
However, the SNfactory uses the wide-field images from the PQ survey to cover thousands of squares degrees 
(Wood-Vasey et~al 2004), like the CRTS. 

In addition to the well-known types of transients listed above, the CRTS discovered a large fraction of OTs
that were either spatially associated with very faint blue galaxies, 
or had no visible source associated with them to the limits of Palomar Quest and SDSS surveys ($>3.5$ magnitudes fainter
than the discovery images).  One well known source of transients associated with galaxies is AGN flares (Totani et~al.
2005).  AGN are known to contaminate SNe surveys (Gal-Yam et~al. 2008, Sand et~al. 2007) and vary on timescales from
hours to years, with outbursts generally being up to $\sim 1$ magnitude (Webb \& Malkan 2000).
We expected to find a number of events from such objects. In contrast, the transients we discovered became many magnitudes
brighter than their hosts over a period of two weeks and were visible for more than a month, consistent with SNe.  We
carried out photometric follow-up of a number of these candidates and found that their colours and decline rates were
indeed consistent with SNe, rather than AGN, CVs or other variables stars (Poznanski et~al. 2002). The most likely
sources of such transients are SNe in very faint host galaxies or associated with intra-cluster stars.

\begin{figure}{
\epsscale{0.8}
\plotone{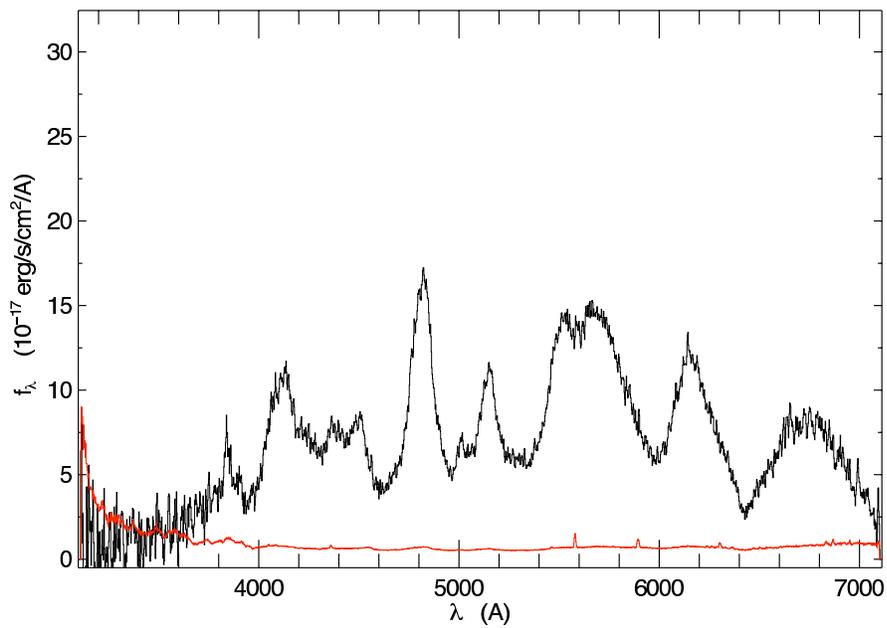}
\caption{\label{spec} The Palomar 200 inch DBMS spectrum of supernova SN 2008ba.}
}
\end{figure}

\placefigure{spec}

To select SNe from among CRTS transients we removed asteroids near stationary points, flare stars and DNe with faint
sources. This was achieved by only selecting transients present at the same location on multiple nights spanning more
that two weeks.  In addition, we required that the objects were at least two magnitudes brighter than any possible host,
and that the transients faded slowly (as expected for SNe). To investigate the accuracy of the classification based on
six months of data we reviewed archival CSS photometry going back more than three years and found no evidence of
contamination by repeated outbursts of variable objects. The SNe candidates are given in Table \ref{tab1}.

\placetable{tab1}

\begin{figure}{
\plottwo{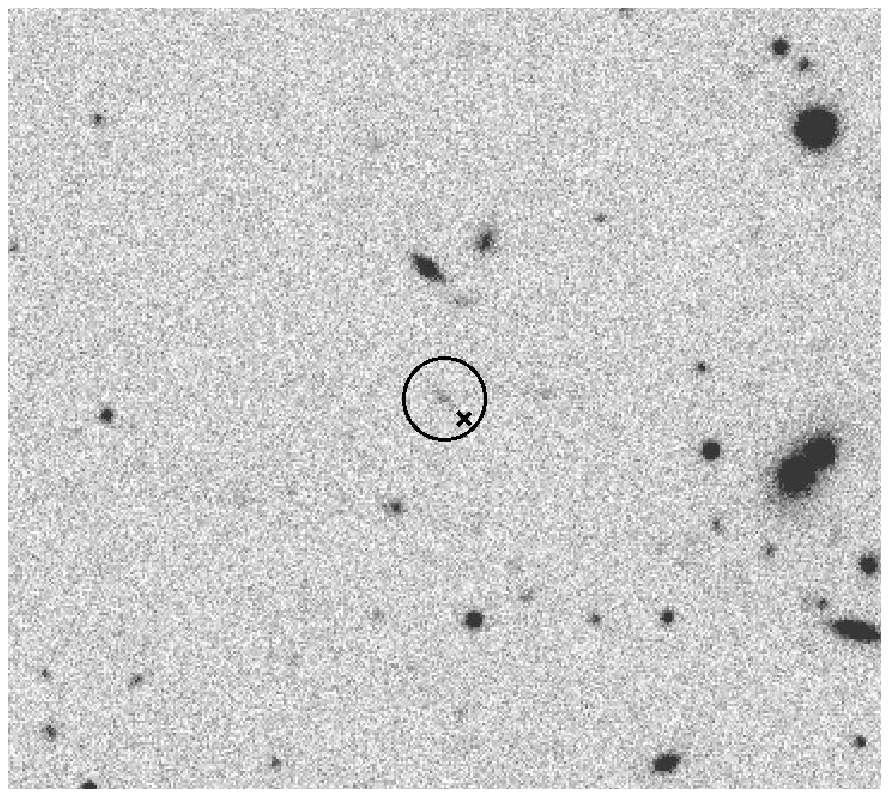}{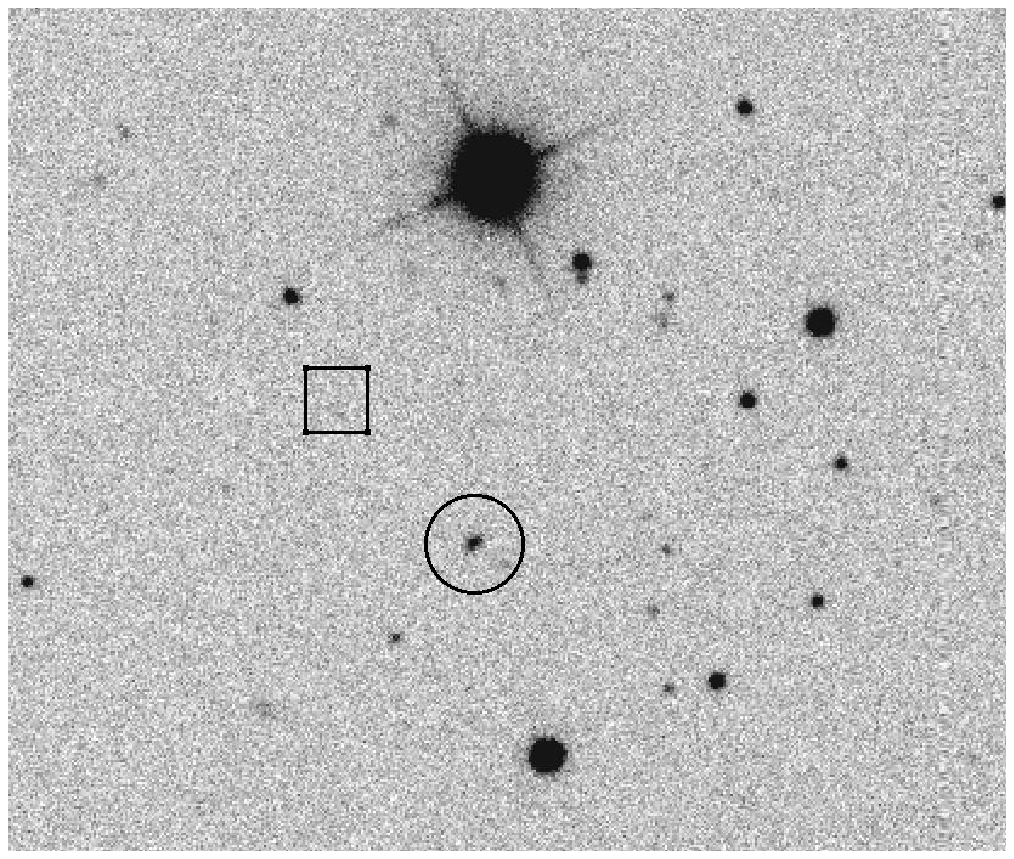}
\caption{\label{img}{\em Left:} SDSS pre-discovery Gunn $r$-band image of the location of SN2008ba.
The circle shows the location of the probable host, while the cross marks the location
of the supernova. {\em Right:} SDSS pre-discovery Gunn $r$-band image of the location of SN2008bb. 
The box shows the location of the SN, while the circle shows the location of the 
nearest SDSS galaxy.
}
}
\end{figure}

\placefigure{img}

Follow-up spectroscopy of two of the CRTS candidates associated with faint galaxies by Djorgovski et~al. (2008c)
revealed a type Ia SN (SN 2008ba, $z \sim 0.03$) and a type II SN (SN 2008bb, $z \sim 0.03$). In Figure \ref{spec}, we
present the spectrum of SN 2008ba. In Figure \ref{img}, we show the locations of the SNe in much deeper SDSS r-band
images.  SN 2008ba is $\sim 4\arcsec$ from an SDSS-identified galaxy with magnitude $m_{r}=21.5$.  There are no other
bright galaxies within $2\arcmin$ of this object.  As the redshift of this SN is 0.03 the host galaxy must be fainter
than $M_{r} \sim -13.9$.  For SN 2008bb there are also no bright ($m_{r} < 19$) galaxies within $2\arcmin$ and no
galaxies with $m_{R} < 23$ within $0.5\arcmin$ (Figure 2, {\em right}).  Assuming the brightest galaxy missed by SDSS to
have $m_{r} \sim 23$, the host would have to be $M_{r} > -12.9$ at the supernova's redshift (0.03). 

\begin{figure}{
\plottwo{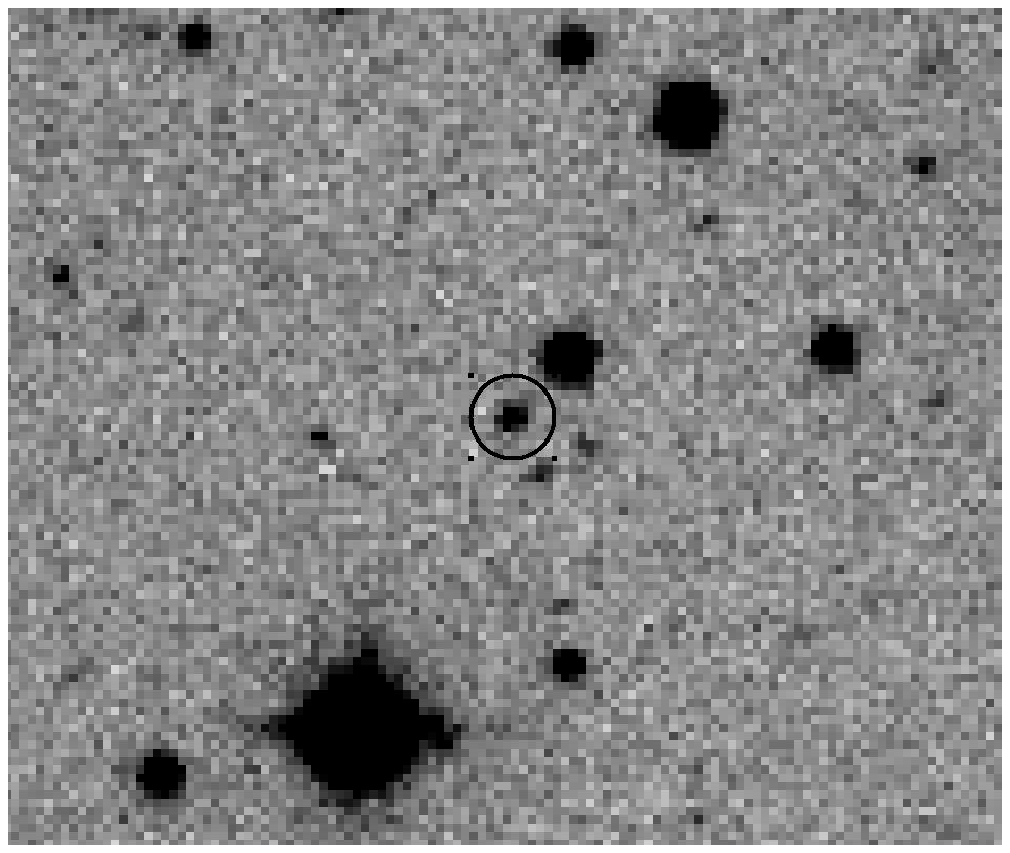}{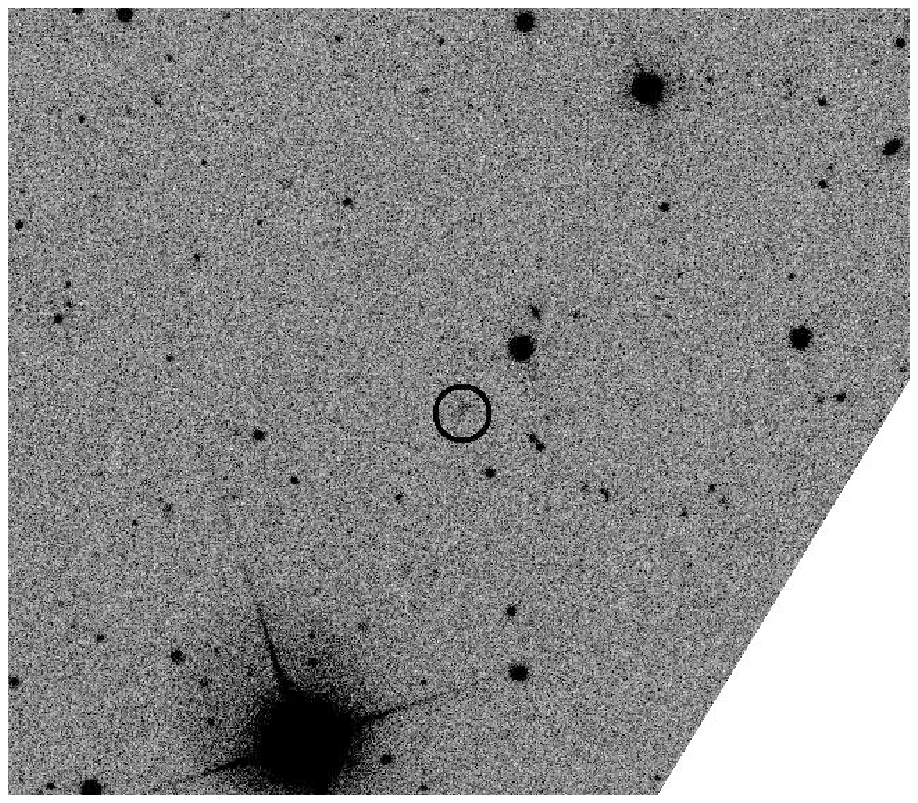}
\caption{\label{712} {\em Left:} 2007-12-30 CSS discovery image, showing the location of OT CSS071230:082550+220041 (circle).
{\em Right:} SDSS Gunn-$r$ image, showing the location of CSS071230:082550+220041 and the corresponding faint galaxy host.
}
}
\end{figure}

\begin{figure}{
\plottwo{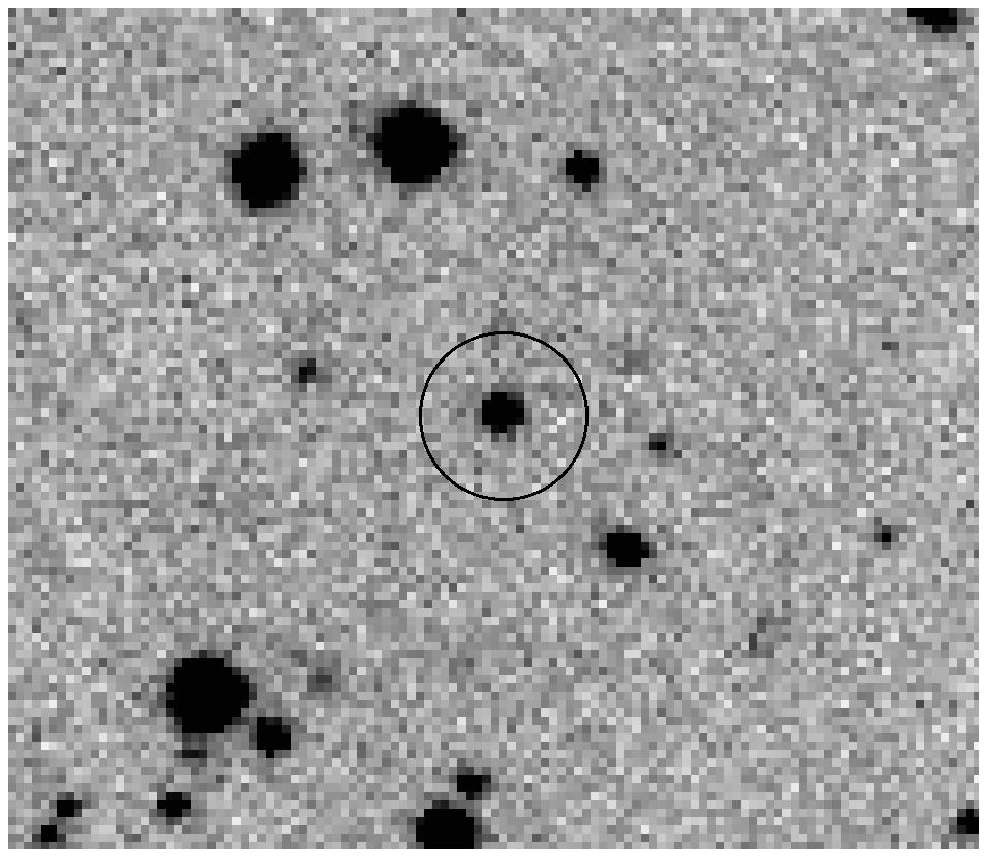}{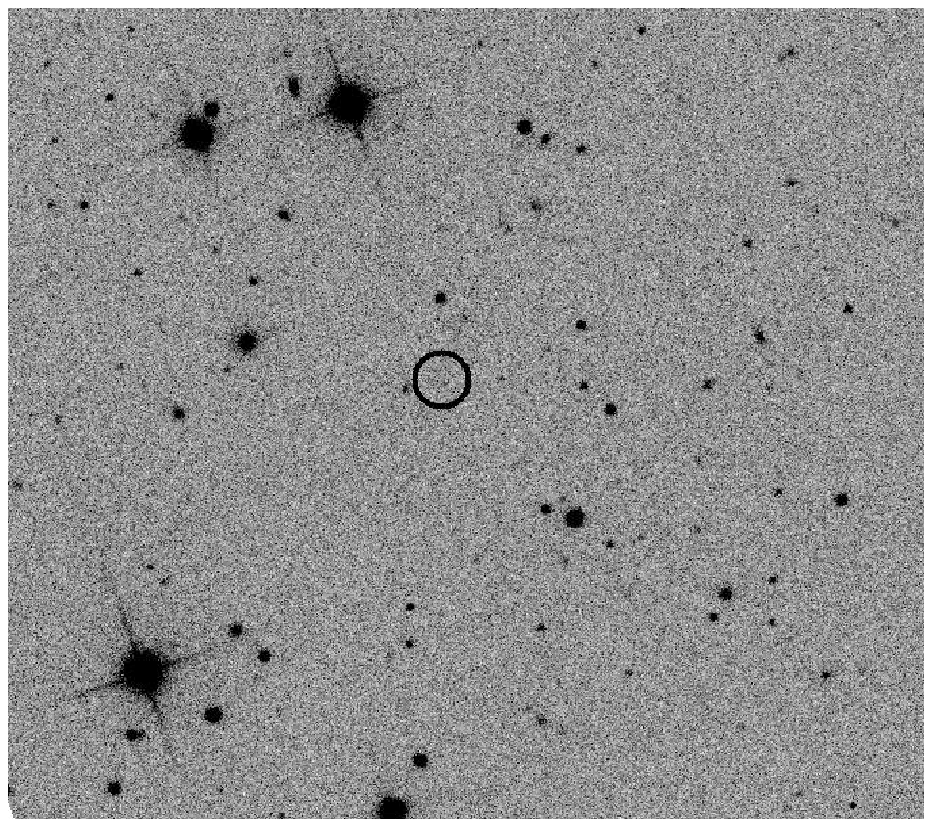}
\caption{\label{802}{\em Left:} 2008-01-01 CSS image, showing the location of OT CSS080219:151457+234110 (circle).
{\em Right:} SDSS Gunn-$r$ image of the same field. At the location of CSS080219:151457+234110, no host galaxy is visible.
}
}
\end{figure}

\begin{figure}{
\plottwo{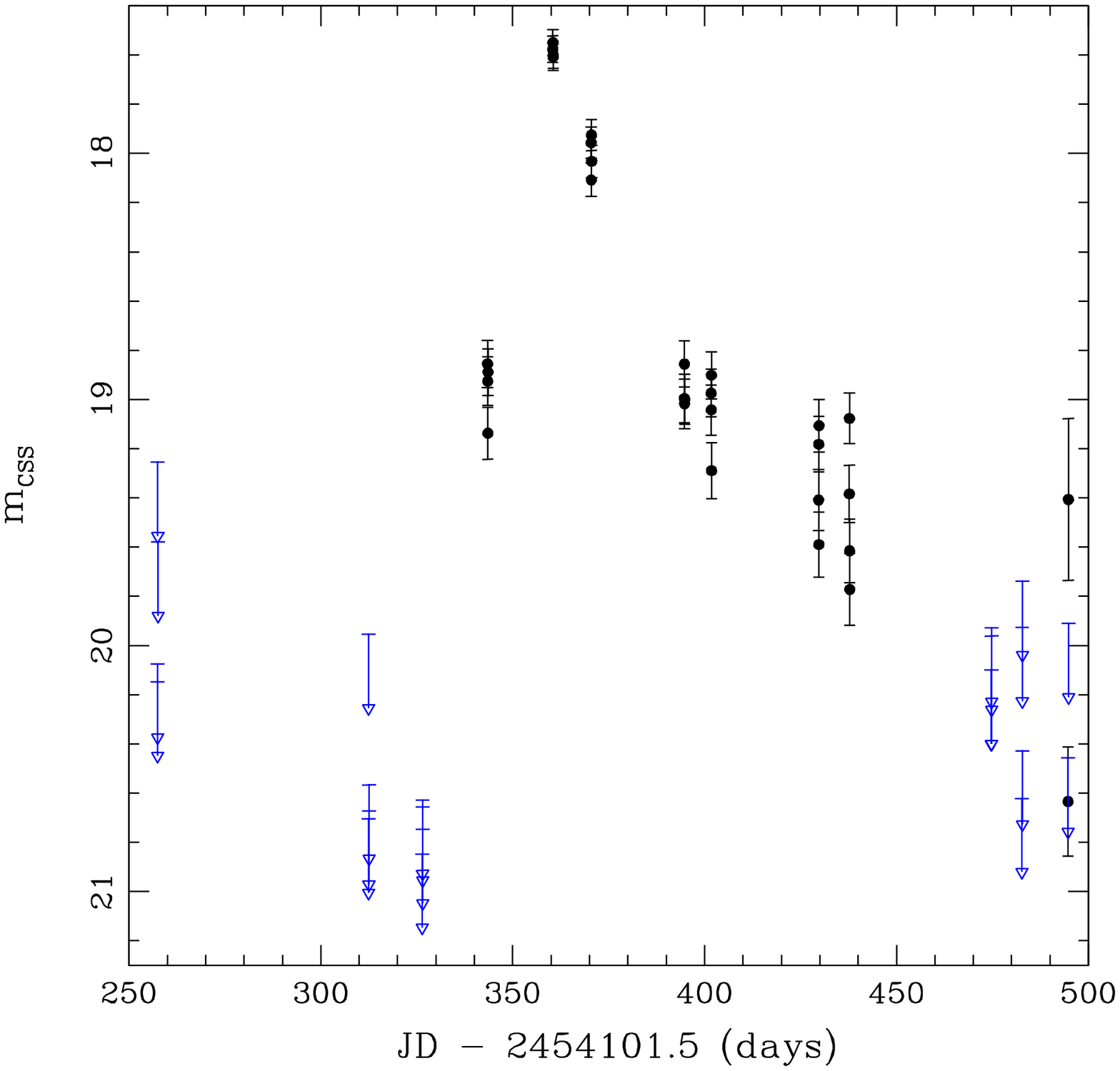}{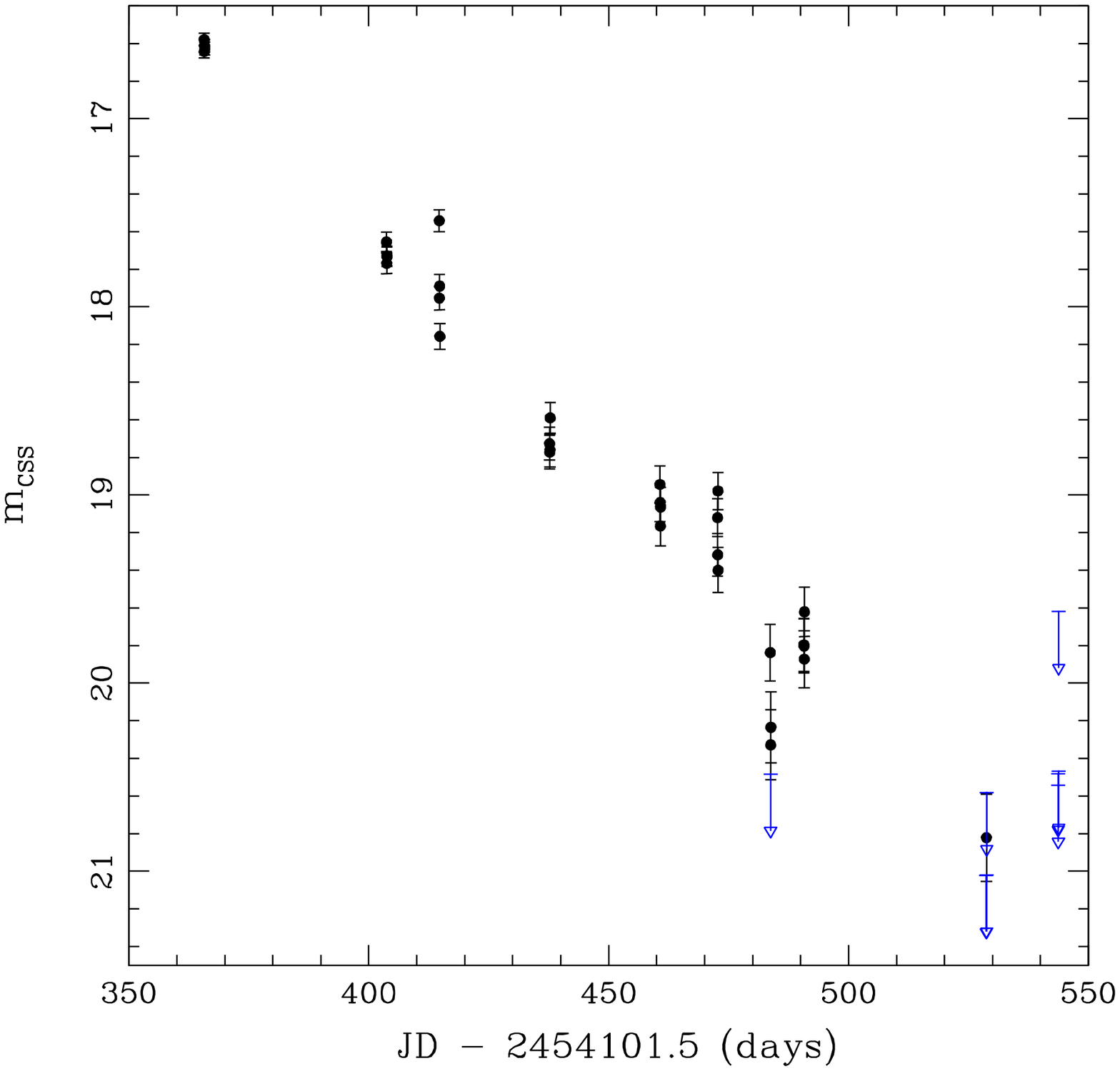}
\caption{\label{LC}
{\em Left:} The CRTS lightcurve of object CSS071230:082550+220041. 
{\em Right:} The CRTS lightcurve of object CSS080219:151457+234110. 
Black points are measurements and blue triangles non-detection upper limits.
}
}
\end{figure}

In addition to the confirmed SNe, we have discovered dozens of similar, long-timescale optical transients.  Seventeen of
these have no sources present in CSS, PQ or SDSS images overlapping the discovery locations (to mags $\sim 21-23$).  The
remaining 24 match with faint galaxies that are $> 2$ magnitudes fainter than the transients associated with them. In
Figure \ref{712}, we present an example of a SN where a faint host galaxy is detected, and in Figure \ref{802}, a SN
where no host is seen to $r \sim 23$.  The lightcurves corresponding to these OTs are given in Figure \ref{LC}.

\placefigure{802}
\placefigure{LC}

Supernovae with faint host galaxies are not unknown. For example, Prieto et al. (2008) studied recent nearby SN and
found the SN Ic, 2007bg, to be associated with a faint ($M_{B} \sim -12$), very metal-poor ($\sim \frac{1}{20}$ solar)
host.  Similarly the Palomar-Quest experiment discovered a type Ic SN (2007nm) with an $m_{R} \sim 22$ host in DPOSS
images and redshift $z=0.04$. At this redshift the host galaxy would have to be $M_{R} > -13.8$.  Another possibility is
that SNe without visible hosts could be associated with intra-cluster stars.  Past searches for such SNe have been
carried out by Sand et al.~(2007) and Gal-Yam et~al. (2008).  As part of the WOOTS survey Gal-Yam et~al. (2003, 2008)
discovered 12 SNe in Abell clusters selected according to $0.06 < z < 0.2$. Two of these were identified as being
intracluster SN-Ia (SN 1998fc and SN 2001al).  They found upper limits to possible dwarf hosts of $\rm M_{r} > -14$ and
$\rm M_{r} > -11.8$, respectively.  Sand et al.~(2007) discovered four photometrically selected intra-cluster SN with
hosts $M_g > -14.3$.  Based on this they found 20\% of stellar mass may reside in intra-cluster stars and thus 20\% of
the SN-Ia parent stellar population could be intergalactic.

Interestingly, Stanek et~al. (2006) and Savaglio et al. (2008) found that the long-duration $\gamma$-ray bursts
associated with SNe are in faint, metal-poor, star-forming dwarf galaxies.
This result was supported by Modjaz et~al. (2008) who found that the galaxies containing type Ic SN that were not
associated with GRBs, were significantly more metal rich than those with associated GRBs. Savaglio et~al. (2008) 
also found that the host galaxies associated with GRBs are low-mass star-forming objects.
The rate of SNe in such low-mass galaxies is expected to be low, as Kauffman et al. (2003) found that only $\sim 13\%$
of the stellar mass in galaxies is found in objects less massive than $10^{9}\, M_{\odot}$. Such galaxies have only tens
of millions of stars in contrast to large galaxies like our own with hundreds of billions of stars. For the hosts of SNe candidates 
not to be visible, most of these must be associated with low stellar-mass galaxies.

\subsubsection{Supernovae with Faint Galaxy Hosts?}

In order to test whether some of our SNe candidates are associated with galaxy clusters we compared 
their locations to Abell clusters (Abell et~al. 1989) and DPOSS clusters (Gal et~al. 2003).
Two of the 48 SNe in the DPOSS catalog region (with Galactic latitude $\rm b > 30\arcdeg$) matched the locations of DPOSS 
clusters and two matched Abell clusters. With only four matches to 12,000 clusters it appears very unlikely
that the SNe are associated with such clusters. The SNfactory, like CRTS, uses wide field images
to find SNe rather than concentrating on bright galaxies like most other nearby SNe surveys. 
Therefore both surveys have similar spatial sensitivity. We compared our results with 95 confirmed SNe from the Supernova 
Factory\footnote{http://snfactory.lbl.gov/snf/open\_access/login.php}. 
Once again only two SNfactory SNe of the 63 SNe with $\rm b > 30\arcdeg$ matched DPOSS clusters and two matched Abell clusters.

To determine whether the galaxies associated with our SNe candidates were indeed low luminosity hosts we calculated the
difference between the peak measured magnitude in CRTS data and the host R-band magnitude.  To provide upper limits to
the host galaxies brightness where no host was visible, the hosts were assumed to reside at the limits of the deepest
observations covering them.  Next, for the SNe with known spectral types (hence known peak absolute magnitude), we added
the CRTS magnitude to the peak expected for the type of SNe.  This provides a rough estimate of the host's absolute
R-band magnitude.  To test the SNe hosts where we do not have known spectral types we carried out the same process and
assumed the SNe were all the brightest, common SN type, Ia. As the actual SNe types should be the same or fainter than
our assumption we obtain an approximate upper limit to the brightness of the associated host galaxies.
For comparison we took the SNfactory SNe and matched them with SDSS DR6 to find their host magnitudes.
Fifty four lie within the region covered by SDSS DR6 and 52 match SDSS objects. Twenty nine of these hosts have spectra 
from which the redshift has been determined. For the other 23 SNe we used the redshift of the SNe given by the 
SNfactory webpage. We then used the SDSS-DR6 r-band magnitudes and redshifts do derive the absolute magnitudes of 
the host galaxies assuming $\rm H_{0} = 72 km/s/Mpc$.

\begin{figure}{
\epsscale{0.7}
\plotone{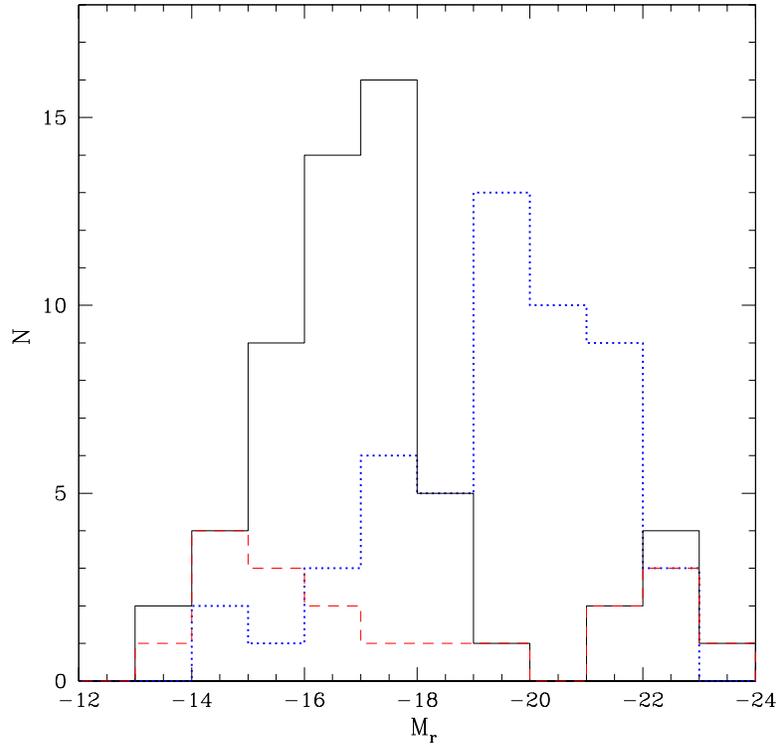}
\caption{\label{snhist}
A histogram of inferred host galaxy magnitudes.
Solid line: all host magnitudes. Dashed line: magnitudes for SNe with known
types. Dotted line: magnitudes from SNe discovered and confirmed by the SNfactory. 
See text for details.
}
}
\end{figure}

\placefigure{snhist}

In Figure \ref{snhist}, we plot histograms of the host galaxy magnitudes, for those with confirmed types and these
combined with candidates (assumed type Ia) as well as those with host from the SNfactory. This figure strongly suggests
that the host galaxies must be intrinsically faint. The spectroscopically-confirmed SN, that provide the most accurate
host galaxy magnitudes, clearly reside in very faint hosts.  The fact that the unconfirmed SN-host distribution peaks at
a brighter magnitude is expected because of our assumption that these SNe were all bright type Ia, rather than a mixture
of types.  Also, it is apparent that the SNfactory finds SNe in intrinsically brighter galaxies.

It is clear that by only selecting transients that are two magnitudes brighter than the sources associated with them, we
will introduce a bias towards discovering SNe in faint galaxies, opposite to SNe surveys that have chosen only to follow
bright galaxies.  However, the CRTS SNe can also be discovered in large galaxies when they rise more than two
magnitudes above the local background. For intermediate magnitude hosts, the incremental increase in brightness
is insufficient for SNe to be detected as separate sources. 
Detection of these would require the use of image subtraction. As the host galaxies discovered by CRTS are generally
much fainter than the SNe they do not add significant flux to follow-up spectra, making confirmation easier.

Two additional possibilities might explain some of the observed SNe candidates.  One possibility is that some of the SNe
discovered may be brighter than SN Ia.  There is increasing evidence for such rare types of bright SNe (Miller et~al.
2008).  An example of such an event is 2006gy, that had a peak luminosity of $\rm M_R \sim -22$ (Smith et
al.~2007).  Based on the small number of discoveries these SNe must be extremely scarce. A second possibility is that some
of the candidates are not SNe, but rather part of a rare group of variable stars. Based on the discovery rate
such hypothetical variable stars would have to be less common than dwarf novae and fainter on average. 
Spectroscopic evidence suggests that many SNe occur in faint hosts, but cannot rule out other possibilities 
for all candidates.

One of the SNe discovered by CRTS (SN 2007sr, Drake et~al. 2007b) resides in NGC 4089/39, the Antennae. This galaxy
is sufficiently nearby ($22\pm 3$ Mpc; Schweizer et~al. 2008) that Cepheid variables can be discovered using the Hubble
Space Telescope.  The HST Key Project (Freedman et~al. 2001) used Cepheid distances to measure the Hubble constant.
However, only eight type Ia SN, occurring between 1937 and 1998, are known in galaxies where Cepheid distances were
measured by the HST Key Project. The combination of a Cepheid distance determination to the Antennae along with well
sampled light curves of SN 2007sr can thus play an important role in independently calibrating the cosmological distance
scale.

\subsection{Cataclysmic Variables}

In this survey we discovered a number of OTs that have abrupt outbursts that lasted for less than a couple of weeks.
These are mainly due to dwarf novae type CVs. Longer outbursts can occur for dwarf novae that undergo
superbursts and standstills. These outbursts can easily be distinguished from classical novae within the Galaxy
because such events have an extreme peak brightness $M_{V} = -7.5$. 
Novae are much more common than SNe, but ten or more magnitudes fainter. Novae are much less 
common than CV outbursts and less luminous than the host galaxies mentioned in the previous section, 
so few are expected in CRTS data.

The brightest outbursts of CVs are closely followed by a large group of astronomers with modest sized telescopes.  For
this reason, most CV follow-up concentrates on objects brighter than V = 17 during outburst.  As LSST images will
saturate at 16th magnitude, the CV outbursts discovered by CRTS are much better suited for follow-up by large numbers of
small telescopes.

\begin{figure}{
\plotone{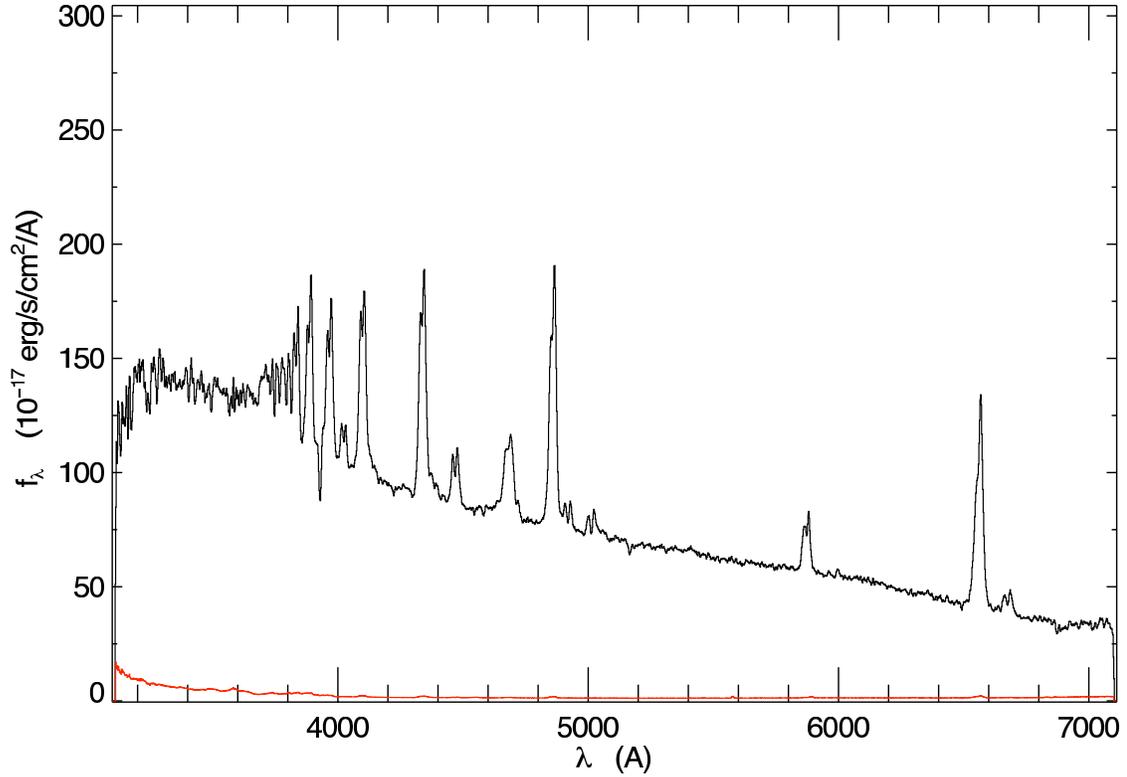}
\caption{\label{CVspec}
The Palomar 200 inch DBMS spectrum of CV eclipsing CV CSS080227:112634-100210.
The Balmer emission lines are clearly seen as well as strong Helium lines.
}
}
\end{figure}

The CRTS transient pipeline has initially been tuned to discover objects brightening by two or more magnitudes between a
catalog magnitude and subsequent observations. This threshold was chosen to discover highly variable sources while
suppressing noise and well known types of low-amplitude variable stars.  However, this threshold makes CRTS ideal for 
the detection of dwarf nova type cataclysmic variables (which typically brighten by 2 to 8 magnitudes). 
In the first six months the CRTS discovered 64 new DNe. All the CV candidates discovered in the CRTS data are 
presented in Table \ref{tab2}. The spectrum of newly discovered eclipsing CV CSS080227:112634-100210
is given in Figure \ref{CVspec}.

\placefigure{CVspec}
\placetable{tab2}

In recent years many bright CVs have been identified in SDSS data from the spectra (Szkody et~al. 2002, 2003, 2004,
2005, 2006, 2007).  To date the SDSS surveys have taken more than 1 million spectra\footnote{http://www.sdss.org}.
However, this is less than $0.5\%$ of the objects observed photometrically and limited to objects brighter than
$i=20.1$.  Thus, most CVs residing in areas covered by SDSS will not have been identified.  Indeed, a large fraction of
the CVs presented in Table \ref{tab2} reside within regions observed by SDSS. As expected, most of these sources clearly
match blue point sources in the SDSS data.  Transient CSS080427:124418+300401 matches SDSS star J124417.89+300401.0 and
has a spectrum that clearly exhibits the strong emission lines associated with CVs but has not previously been
identified as a CV by the SDSS team.  However, many of the DNe we discovered (without spectra) are noted as SDSS QSO
targets. These were excluded in the main SDSS QSO survey (Richards et~al. 2002), but were later included in the faint
SDSS QSO extension survey.  Clearly the reason that these are QSO candidates is simply that CVs and QSOs overlap in the
colour-colour space of SDSS images. Since CVs are strong UV sources, we matched DNe candidates within the far
ultraviolet and near ultraviolet observations of GALEX. White dwarfs and QSOs also overlap when the GALEX UV
observations are combined with SDSS observations (Bianchi et~al. 2007).  The detection of these objects in GALEX FUV and
NUV data provides additional evidence that these are CVs.  At the current CRTS CV discovery rate, using archival CSS
data we expect to find $\sim 300$ new CVs. In comparison, Szkody et~al. (2002; 2003; 2004; 2005; 2006; 2007) discovered
177 new CVs in 6 years of SDSS data, while Rykoff et~al. (2005) discovered four CVs in ROTSE-III data taken between
September 2003 and March 2005.

In this survey we selected CVs by matching the object locations with SDSS-DR6 data (Adelman-McCarthy et~al. 2008). Most
CVs overlapping SDSS clearly match blue objects marked as stars in SDSS observations. For those CV candidates not
covered by SDSS-DR6 data we checked DSS1, DSS2 and Palomar-Quest Johnson and Gunn images looking for prior outbursts.
For CV candidates without SDSS spectra, matches with blue sources, or evidence of prior outbursts in archival images, we
required that the CRTS detection was made at the same location on more than one night. We also required that there be
additional photometry showing that the object had returned to quiescence within a few weeks. In a small number of cases
the CV locations were covered by DSS as well as SDSS or PQ data, yet no sources were present in any of the past images.
These objects were classified based on their lightcurves.  

Many types of CV systems exist.  UMa type CVs exhibit superbursts of up to $\sim 8$ magnitudes.  Similarly, WZ Sge type
CVs have large outbursts. However, the interval between bursts can be decades.  In order to alert astronomers of
outbursts we initially posted alerts for new discoveries in Astronomer's Telegrams
(ATels)\footnote{http://www.astronomerstelegram.org}. As the classification process progressed we posted candidates to
the CVNet group\footnote{http://tech.groups.yahoo.com/group/cvnet-discussion/}.  However, many of the CVs were also
independently followed by members of VSNET\footnote{http://ooruri.kusastro.kyoto-u.ac.jp/pipermail/vsnet-alert/} within
hours based on the real-time alerts posted on the VOEventNet webpages.
Subsequent observations of some of the CVs by members of these groups showed the presence of 
superhumps (seen in UMa type systems). The timing of these humps gives the orbital period of the systems. 
In addition, some of the CVs we discovered were later found to exhibit outbursts in archived ASAS-3, 
DSS and PQ images.

Recently, Rau et~al. (2007) undertook follow up observations of four optical transients discovered in two wide field
surveys. Of these one was discovered to be an asteroid and the other three were cataclysmic variables. In addition they 
determined limits on the number of faint CVs expected at high Galactic latitude. Based on their model they predicted
that CVs with quiescent magnitudes fainter than R=22 should not be discovered at Galactic latitude $|b| > 45\arcdeg$.  
In our survey we have so far found 2 of the 16 CVs fainter than $R=22$ at galactic latitudes above this.

\subsection{Eclipses}

Apart from searches for objects that newly appeared in a sequence of CSS images, or varied on average from catalog magnitudes, we
searched for intra-sequence variability. That is, objects with real brightness variations within the 10 minute
span between each image in a sequence of four.  This search revealed a number of candidates that dropped by 
more than a magnitude from the baseline brightness between observations. 

These transients are of some interest as the eclipses observed for stellar binary systems are typically much less 
than a magnitude.  Larger eclipses require that the eclipsed star be brighter and smaller than the eclipsing star.
As brighter stars are generally larger, they are not fully eclipsed by their companions.
One of the exceptions to this situation occurs for white dwarf stars.
These objects are much smaller than main-sequence stars of a similar bolometric luminosity.
Thus, large eclipses can occur when they are eclipsed by late type dwarf stars.
We found SDSS spectra corresponding to two of these objects which clearly show the presence of white dwarf-M dwarf
binary systems.  In addition, a couple of objects were found to exhibit outbursts and flickering. These are eclipsing CV
systems where unlike white dwarf binary systems the drop can be due to the eclipse of an accretion hot spot (Brady et~al.
2008).

Eclipses of stars are an effect that can be used to constrain the radii of the eclipsing objects.
Steinfadt et al.~(2008) recently discovered a partially eclipsing white dwarf (SDSS J143547.87+373338.5)
and used it to determine the characteristics of the M-dwarf.
When secondary eclipses are observed in partially eclipsing binary systems, it is possible to constrain their 
orbital inclination, eccentricity and orientation. When no secondary eclipse is 
seen, the cause of this may be a combination of inclination, orientation and orbital eccentricity, 
or the low relative luminosity of the eclipsed star. 
The existence of full eclipses provides a strong constraint on the inclination of such systems, wheras the
depth and length of the eclipse constrains the relative effective temperatures and the radii.
Variations in luminosity during a full eclipse can also be used to map surface features of an 
eclipsed star, such as spots. 
For white dwarf binaries, 
more fully eclipsing systems are expected than partially eclipsing systems, simply because 
radii of other types of stars are $>10$ larger than white dwarfs (L\'opez-Morales et~al. 2007).

\begin{figure}{
\plottwo{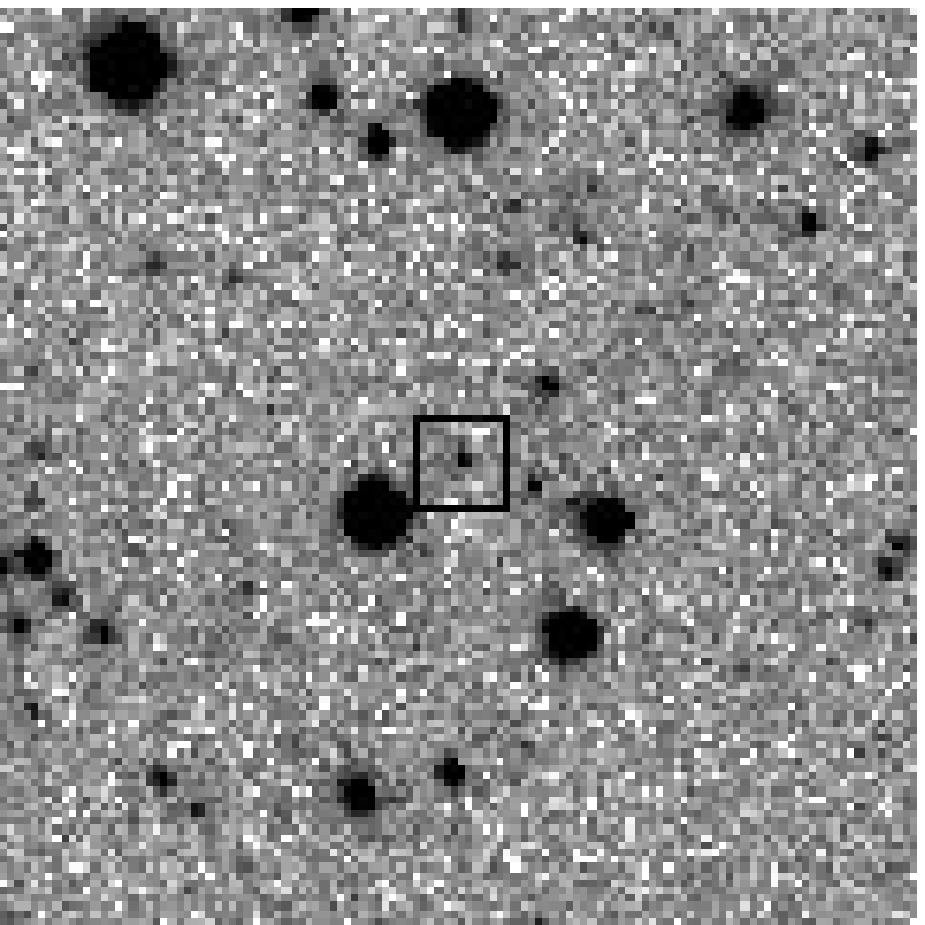}{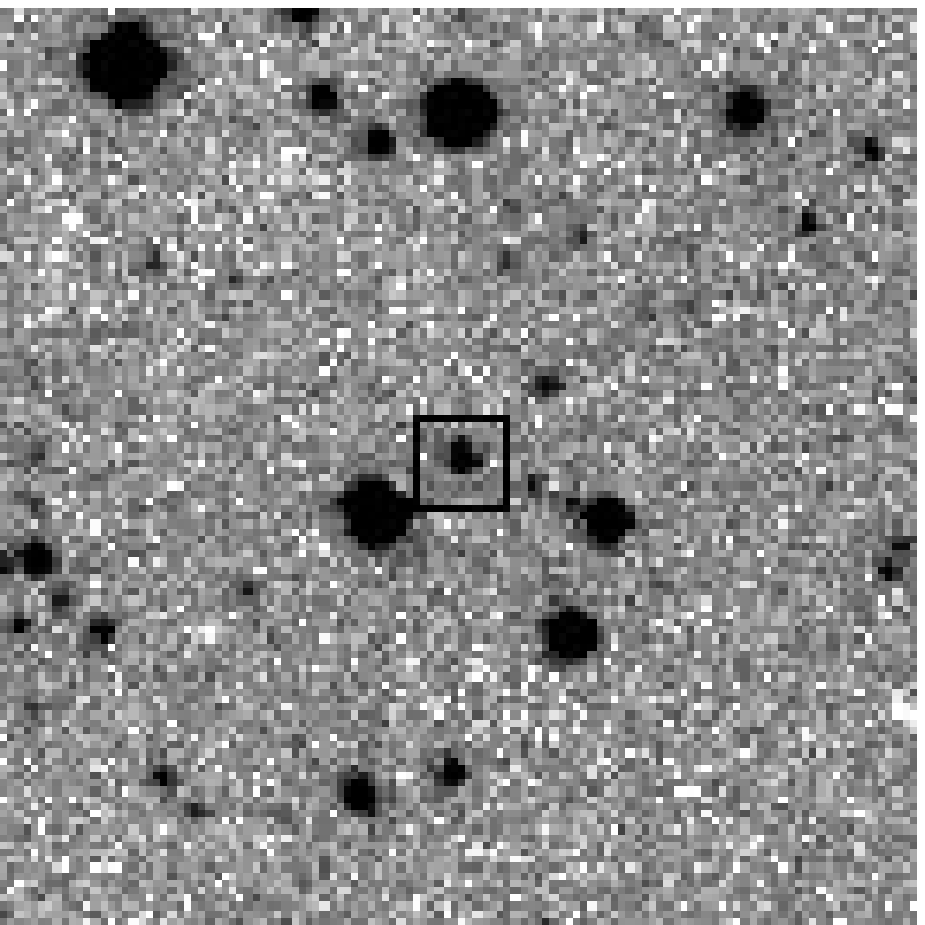}
\caption{\label{WDa} CSS images of object $\rm CSS080502$:$090812$+$060421$ in and out of eclipse.
The baseline CSS magnitude is $17.0\pm0.1$, while the individual CSS image magnitudes are $18.72 \pm 0.09$, $18.91 +/-0.10$, $17.23+/-0.04$, $17.15\pm0.04$.
This system matches SDSS object J090812.04+060421.2 with mags $\rm u=17.34$, $\rm g=17.08$, $\rm r=17.28$, $\rm i=17.08$,$\rm z=16.69$.
This object was found to be a DA white dwarf of subtype 2.8, plus an M3 dwarf by Silvestri, et~al. (2006) also 
matches 2MASS09081205+0604211 (mags: $\rm J=15.5$, $\rm H=14.9$, $\rm K=14.7$) and GALEX source J090812.0+060420
($\rm FUV = 17.1$ and $\rm NUV =17.2$).
}
}
\end{figure}

\placefigure{WDa}

\begin{figure}{
\plottwo{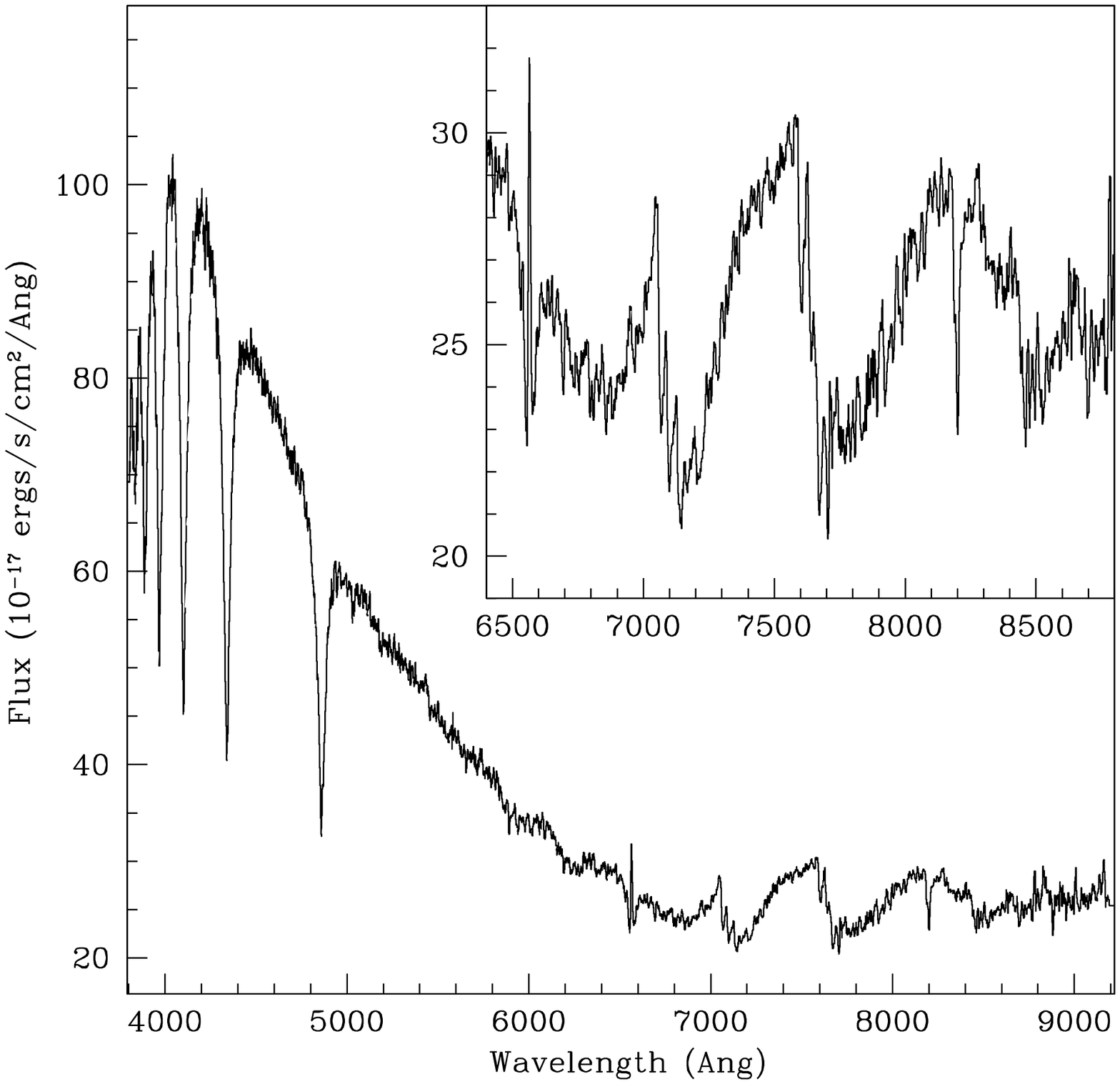}{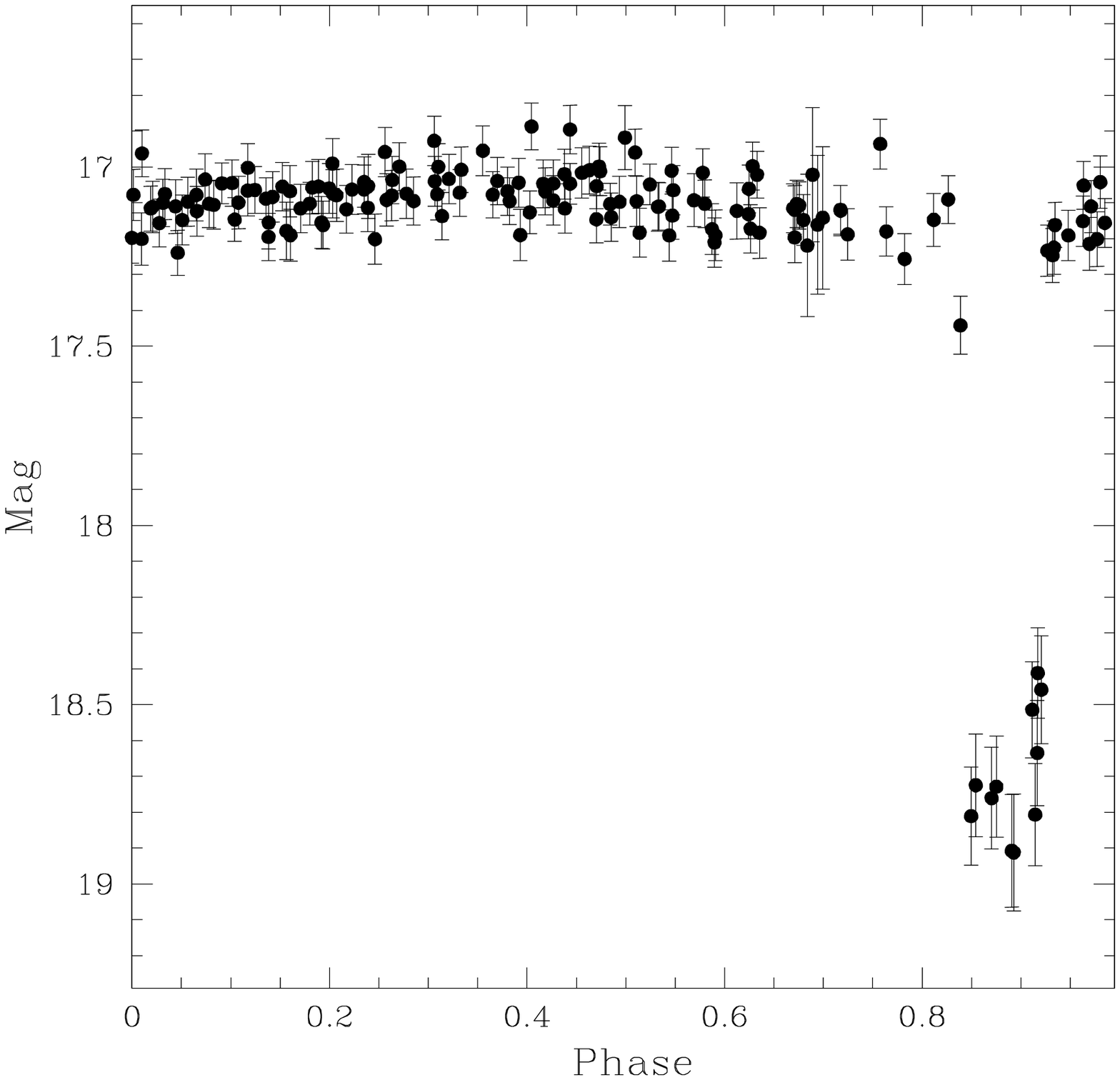}
\caption{\label{WDb}
{\em Left}: The 5-pixel smoothed SDSS spectrum of $\rm CSS080502$:$090812$+$060421$.
The inset shows an expanded view of the M-dwarf TiO features.
{\em Right}: the phased lightcurve of $\rm CSS080502$:$090812$+$060421$ with period $3.58652\pm 0.00002$ hours.
}
}
\end{figure}

Two clear WD-MD binary systems with eclipses $>1.7$ magnitudes in depth were discovered.
In Figure \ref{WDa} the WD-MD binary system CSS080502:090812+060421 is presented.
This is a known WD-MD binary (Silvestri et~al. 2006), while CSS080408:142355+240925 is 
a newly discovered system.
In both cases the M-dwarf can be seen in the SDSS spectra and its flux is detected in 2MASS data. 
The 2MASS magnitudes and colours consistent with M-dwarf companions (Wachter et~al. 2003).
Using three years of archival CSS data we determined the period of CSS080502:090812+060421 to be 3.58652 hrs. 
In Figure \ref{WDb}, we present its spectrum and phased lightcurve.  The shape of the lightcurve outside eclipse
suggests that the M-dwarf is tidally distorted (Drake 2003). The presence of weak $H_{\alpha}$ emission in the spectrum suggests 
that this is probably a pre-CV system.

\placefigure{WDb}

\subsection{UV Ceti Variables}

One goal of the CRTS is the discovery of transients that may brighten and fade on timescales of
minutes. In the course of the analysis we discovered a number of optical transients by searching for 
significant variations ($> 2$ mags) in two images within the sequence of four images (taken over a 
span of 30 minutes).
In some cases the source of the events could be seen in the co-added CSS catalog images, although in other cases there
was no apparent source. As expected, inspecting SDSS-DR6 images at the locations of these events revealed clear matches
with faint M-dwarf stars. Thirteen of these events were clearly due to flare stars.  An example of this is the
``Lynx OT'', the first announced CSS transient (Christensen 2004; Djorgovski, Gal-Yam, \& Price 2004). These events
are simply flares exhibited by UV Ceti variables.
These flares were discovered at Galactic latitudes ranging from $-60\arcdeg < b < 70\arcdeg$.
The largest observed flare was 5.5 magnitudes brighter than an associated SDSS source.
Aside from stellar flares associated with M-dwarfs, there were transients showing repeated large variations. These
variations are thus possibly related to active M-dwarf stars. The CRTS is in an excellent position to resolve the 
frequency and general nature of UV Ceti flares, as the timescales of flaring events are tens of minutes,
and the CSS survey takes four images separated by 30 minutes.

Past optical transient surveys have also discovered flare stars. For instance, the DLS discovered three faint short
timescale ($\sim 1000$ s) transients during the course of their three year survey (Becker et~al. 2004). One of these
transients was spectroscopically identified as a flaring dM4 star. The other two were later identified as flare stars by
Kulkarni \& Rau (2006). These authors noted the prevalence of flare star detections in past transient surveys.
Additionally, they noted that the presence of a large ``fog'' of such stars would
hamper future efforts to discover fast transients. 
From the transients discovered in this survey it is clear that separation of flare events from other fast transients 
can easily be achieved with observations taken on timescales of flaring events.  That is to say, there are distinctive
brightness variations between images. In this study, classification is further aided by the presence of much deeper
photometry (from PQ, SDSS-DR6) where the M-dwarf stars can be clearly identified. Future large surveys such as 
LSST will go much deeper than SDSS (Ivezic et~al. 2008). These deep experiments will have difficulty detecting the 
faint flaring sources without a significant amount of deeper follow-up. However, it is also clear from these results 
that large flaring events lasting more than ten minutes are much less common than large outbursts of DNe. 

Recently, Welsh et~al. (2007) found 52 flares associated with 49 flare stars in GALEX data. They discovered two kinds 
of flare events, namely events lasting $< 500$s with quasi-exponential decays and events lasting longer than 500s with
secondary emission peaks. Thirty nine of the GALEX flares were greater than two magnitudes. This is three times the 
number of CRTS discoveries, yet the GALEX coverage surveyed (in square degree seconds) is only 20\% of the CRTS data 
analysed here. However, 27 of the GALEX flares lasted $< 500$ seconds. Such events would appear as a single point and 
thus not be discovered in CRTS data. In addition, GALEX's observations reach much fainter magnitudes 
($\rm m_{NUV} \sim 21$) than CRTS.

There is some evidence that the flaring activity of some late M-dwarf stars has been associated with the presence of
stellar companions (Schloz 2004).  One of the transients we discovered (CSS080118:112149-131310) was a $> 3.5$ magnitude
flare on a nearby high proper motion star, LHS-2397a.  This star has a proper motion of $0.509\arcsec/yr$ and was
serendipitously observed spectroscopically during a flare by Bessell (1991).
LHS-2397a is in fact a binary system where the primary star is an M8 type star (LHS-2397aA) and the 
companion (LHS-2397aB), a L7.5 Brown Dwarf orbiting within 4 AU (Freed et al. 2003).

\subsection{Asteroids and Short Timescale Optical Transients}

The only interest of the CRTS is the discovery of transients beyond our Solar system. Moving transients are 
discovered by the CSS in their search for NEOs. However, to detect faint transients we have to allow for moderate
astrometric uncertainties between detections.  Thus, our transients selection is limited to objects that move an average
of $\rm < 0.1\arcsec/min$ between observations. In order to determine how asteroids near stationary points might
contaminate extra-solar transient detections we investigated the distribution of known asteroid apparent motions.  We
chose the night of October 21st 2006 which contains 21,492 known asteroids in the CSS fields.  Of these objects 2.2\%
were predicted to have apparent motion $\rm < 0.1\arcsec/min$ when observed.  In our processing we remove asteroids by
comparing the locations of transients with all known asteroids determined from MPC ephemerides.  Furthermore, our
detections are limited to a magnitude of $\sim 20$ where a large fraction of asteroids are known.  Nevertheless, a small
number of OTs were seen in only one sequence of images and many of these are probably unknown asteroids.  As objects
such as TNOs have motions of a few arc-seconds per hour, very bright TNOs could be found in CSS data. However, as there
are only $\sim10$ known TNOs brighter than $V \sim 20$, and extensive surveys have been carried out to find these, it is 
unlikely that new TNO will be discovered in CSS data.

The main LSST survey is planning to take pairs of observations in a single filter separated by 15 seconds. A one
magnitude change of a typical flare star over 500 seconds corresponds to a 0.03 mag change during this time. This will
be undetectable for most stars observed by LSST, particularly at the near faint limit where most late M-dwarfs can be 
expected. In addition to stellar flares, over a 15 second time span, faint, unknown asteroids will appear stationary. For example,
an asteroid with an apparent motion of $0.1\arcsec/min$ will move only $0.025\arcsec$ between observations. 
The LSST intends to make two pairs of observations for each field on the same night. This cadence enables asteroid 
detections to be linked together to form arcs that can in turn be linked to form orbits. If the same field was 
observed again after 500 seconds (in order to veto UV Ceti flares), the change in brightness of a flare star would 
be significant. However, the asteroid in our example would only have moved $0.8 \arcsec$. This seems very unlikely 
to provide an arc that can be uniquely linked with a detection 24 hours later.
Deep photometric or spectroscopic follow-up observations could be used to determine the nature of individual rapid
transients discovered by LSST. However, with a conservative estimate of hundreds of flare events every night, a significant
amount of time will be required to follow even a fraction of these objects. 
If the LSST was to take a third observation on the same night, flare stars and moving objects could be clearly 
separated. 

Recently, Horesh et al.~(2008) used overlaps between two SDSS scans to select supernova candidates. 
They removed asteroids by imposing an $\rm 0.16 \arcsec/min$ limit on motion. Using the CSS data
we found $\sim 6\%$ of the known asteroids had motions less than this limit. Although these authors 
removed known asteroids, two thirds of asteroids at the magnitudes of their SNe candidates 
($r \sim 20-21.5$) are unknown (Ivezic et al.~2002). Clearly care will be required when selecting 
short-timescale OTs.

\subsection{Blazars and other variable sources}

Although the main types of transients we have discovered so far are supernovae, cataclysmic variables and UV Ceti stars,
there are other clear populations of transients.  These include nearby high proper motion stars, Blazars and 
highly-variable stars (such as Miras).  

Two possible Blazars were discovered in CSS data and spectroscopically followed, CSS080506:120952+181007 and CSS080409:154159+235603 (Mahabal et al.~2008b).
CSS080409:154159+235603 was fainter than magnitude 20.5 on January 1st 2008 and when observed on February 9th was mag
18.8. It has been approximately the same brightness since discovery, suggesting the object is being fueled by a significant 
source of energy.  CSS080506:120952+181007 was discovered at magnitude 15.8 on May 6th 2008 and matches the flat spectrum 
radio source MG1 J120953+1809 (Helmboldt et al.~2007). Based on the lightcurves of these objects, we have selected
additional transients with similar abrupt and irregular lightcurves.  All these objects are presented in Table \ref{tab3}.

\placetable{tab3}

CSS080426:165347+164950 matches a radio source seen in NVSS data and appears to vary by one magnitude in as little as 
20 minutes. CSS080306:141549+090354 corresponds to the flat-spectrum radio source RGB J1415+090 (Jackson et al. 2007) and 
appears to rapidly vary between magnitude 18.4 and 20.4. CSS080208:120722+250650 matches an uncatalogued source seen 
in SDSS images. Based on nearby sources we estimate that before detection it was magnitude $m_r \sim 23$.  This object 
faded from magnitude $\sim 19.2$ to $\sim 20.2$ over a period of 140 days. Detailed follow-up of these objects
is required to understand their true nature.

\section{Summary and Discussion}

We have presented a few of the types of optical transients found in the first six months of CRTS analysis of CSS
observations. With the observational parameters of this survey the transient discoveries are dominated by two types of
objects, DNe and SNe.  Apart from Miras, few of the many kinds of highly variable stars were discovered.
We believe the main reasons for this are our high detection threshold (2 mags), use of catalogs based on co-added images
(where cyclic variability on short timescales are averaged out), and the natural limit to the number of stars observed
by only observing fields with $|b| > 10\arcdeg$.

While photometric follow-up was taken for many of the OTs in the first six months (Mahabal et al. 2008c), spectroscopic
follow-up was carried out to classify a small number of the candidates. Clearly, one of the most important next steps is
that of classifying these and other such transients, such that the expensive spectroscopic resources are used to a
minimum. Given that surveys tend to provide a small number of data points, this requires optimizing photometric
follow-up in combination with probabilistic classification techniques. We have been making steady progress to tackle
this hard problem using advanced mathematical and statistical methodologies (see e.g. Mahabal et al. 2008d).  As the
survey progresses we will also optimize transient discovery by employing machine learning techniques (Borne 2008). This
will enable unsupervised rapid follow-up to be carried out on the most interesting objects.

The CRTS has a firm commitment to making discoveries and the associated meta-data public as quickly as possible. In this
way we hope all aspects and types of OTs can be explored. Many of the CRTS OTs were posted in ATels and a number of
these were independently observed after these announcements. We chose to announce most of the obvious SNe with bright
hosts in CBETs and these were usually spectroscopically followed and thus confirmed. Although significant follow-up
would be required to characterize all the OTS discovered by CRTS, this is possible as the SDSS-II Supernova
Survey acquired $\sim180$ nights of spectroscopic time on 2.4 to 11 meter telescopes, to follow SNe found during
$\sim70$ nights of imaging (Frieman et~al.  2008).  All OTs will continue to be made publicly available from 
VOEventNet and those events that appear the most interesting will be announced in ATels and CBETs.
In addition, we will be using the two additional CSS survey telescopes to search for transients.  These will allow 
us to find similar numbers of transients in the Southern sky as well as fainter transients in the North.

In conclusion, while there is little doubt that many of the ``major advances in our understanding of the Universe 
have historically arisen by improvements in our ability to see'' (Ivezic et~al. 2008),  
we believe that significant advances have, and will continue to be made by also changing how and what is observed. 
The upcoming next-generation of transient surveys (PanSTARRs and LSST) promise a unique ability to discover rare 
types of optical transients in the faint time-domain sky. However, it is clear that the full characterization 
of many kinds of transients will require follow-up observations and these are most easily achieved when the 
transients are both accessible to the entire astronomical community, and bright enough to be followed by the 
large numbers of existing small telescopes. 

\begin{acknowledgements}

   We would like to thank J. Greaves and members of the CVNet for the contributions and discussions about 
   cataclysmic variables.
   
   This work is supported by the National Science Foundation under Grant No. CNS-0540369.
   The CSS survey is funded by the National Aeronautics and Space Administration under Grant No. NNG05GF22G 
   issued through the Science Mission Directorate Near-Earth Objects Observations Program.
   The Palomar-Quest digital sky survey is a collaborative venture between California Institute of Technology (Caltech) and
   Yale University. The data are obtained at the Samuel Oschin telescope at Palomar Observatory, and processed at the
   Center for Advanced Computing Research (CACR) at Caltech, using techniques developed in part for the U.S. National
   Virtual Observatory (NVO). The PQ survey is supported by the U.S. National Science Foundation under Grants AST-0407448 
   and AST-0407297.  Support for M.C. is provided by Proyecto Basal PFB-06/2007, by FONDAP Centro de Astrof\'{i}sica
   15010003, and by a John Simon Guggenheim Memorial Foundation Fellowship.

  This research has made use of the SIMBAD database, operated at CDS, Strasbourg, France. This research has made use of
  the NASA/IPAC Infrared Science Archive and NASA/IPAC Extragalactic Database (NED), which are operated by the Jet
  Propulsion Laboratory, California Institute of Technology, under contract with the National Aeronautics and Space
  Administration.
  GALEX is a NASA Small Explorer, operated for NASA by California Institute of technology under NASA contract NAS-98034.  
  Funding for the SDSS and SDSS-II has been provided by the Alfred P. Sloan Foundation, the Participating Institutions,
  the National Science Foundation, the U.S. Department of Energy, the National Aeronautics and Space Administration, the
  Japanese Monbukagakusho, the Max Planck Society, and the Higher Education Funding Council for England. The SDSS Web
  Site is http://www.sdss.org/.

\end{acknowledgements}

\newpage

\clearpage
\begin{deluxetable}{lccccccc}
\tablecaption{Confirmed and Candidate Supernovae.\label{tab1}}
\scriptsize
\tablewidth{0pt}
\tablehead{\colhead{ID} & \colhead{RA} & \colhead{Dec}  & \colhead{Date} & \colhead{Mag} & \colhead{$Mag_{H}$} & \colhead{Designation} & \colhead{Type}\\
& h.m.s & $\arcdeg$ $\arcmin$ $\arcsec$  & $MJD$ }
\startdata
CSS080514:145613+185115 & 14:56:13.27 & +18:51:15.4 & 54600.18572 & 18.4 & 23 & 2008dk & Ia\tablenotemark{$\dagger$} \nl
CSS080508:112311+341251 & 11:23:10.90 & +34:12:51.0 & 54594.27706 & 18.4 & 16.4 & 2008cz & Ia \nl
CSS080505:160625+100521 & 16:06:25.06 & +10:05:21.4 & 54591.37061 & 18.8 & 21.8 & 2008df &  \nodata \nl 
CSS080505:155415+105825 & 15:54:15.15 & +10:58:25.0 & 54591.37006 & 16.7 & 16.8 & 2008cg & Ia \nl  
CSS080429:132846+251644 & 13:28:46.47 & +25:16:44.2 & 54585.25061 & 18.6 & 21.2 & \nodata & \nodata \nl
CSS080427:152242+302208 & 15:22:42.30 & +30:22:08.0 & 54583.39125 & 18.5 & 21.7 & 2008ck & Ia \nl
CSS080426:084328+302346 & 08:43:27.86 & +30:23:45.6 & 54582.15121 & 18.2 & $>$23 & \nodata & \nodata \nl  
CSS080417:155525-094147 & 15:55:24.97 & -09:41:47.1 & 54573.37332 & 16.7 & 21\tablenotemark{$\star$} & 2008de & II\nl  
CSS080415:122958+283538 & 12:29:57.71 & +28:35:38.1 & 54571.16946 & 19.5 & $>$23 & \nodata & \nodata\nl
CSS080414:134142-162022 & 13:41:42.18 & -16:20:21.8 & 54570.31002 & 18.4 & 21\tablenotemark{$\star$} & \nodata & \nodata\nl
CSS080409:170404+213542 & 17:04:03.60 & +21:35:42.0 & 54565.34357 & 17.7 & 20.1 & 2008dd & \nodata\nl
CSS080406:164729+091826 & 16:47:29.45 & +09:18:25.8 & 54562.46315 & 18.1 & 21\tablenotemark{$\star$} & 2008dc & Ib/c\nl
CSS080406:122515+064526 & 12:25:15.03 & +06:45:26.2 & 54562.35553 & 18.9 & $>$23 & \nodata & \nodata \nl
CSS080405:155205-000627 & 15:52:04.82 & -00:06:27.0 & 54561.46586 & 18.7 & 21.2 & \nodata & \nodata \nl
CSS080404:130414-101913 & 13:04:14.11 & -10:19:12.9 & 54560.34851 & 15.1 & 11.8 & 2008aw & II\nl
CSS080404:125030-105201 & 12:50:30.36 & -10:52:01.3 & 54560.34718 & 15.9 & 13.1 & 2008aq & IIb\nl  
CSS080403:123041+413816 & 12:30:40.80 & +41:38:16.1 & 54559.20966 & 13.3 & 9.4 & 2008ax & IIn\nl
CSS080401:092227+282421 & 09:22:27.05 & +28:24:20.8 & 54557.22644 & 18.7 & 23\tablenotemark{$\star$} & \nodata & \nodata\nl
CSS080330:121539+033613 & 12:15:39.10 & +03:36:12.9 & 54555.30168 & 18.8 & 20.6 & \nodata & \nodata\nl
CSS080330:124441+051438 & 12:44:41.47 & +05:14:37.6 & 54555.30276 & 19.2 & 21.4 & \nodata & \nodata\nl
CSS080329:130259+103027 & 13:02:58.75 & +10:30:27.0 & 54554.30700 & 17.2 & 15.8 & 2008bm & IIn\nl
CSS080313:132560-020027 & 13:25:59.69 & -02:00:27.1 & 54538.36493 & 18.6 & 22.1 & \nodata & \nodata\nl
CSS080312:140609+242013 & 14:06:08.61 & +24:20:12.5 & 54537.39966 & 19.4 & $>$23 & \nodata & \nodata\nl  
CSS080312:102245+021753 & 10:22:44.65 & +02:17:52.5 & 54537.27925 & 18.9 & 21.7 & \nodata & \nodata\nl
CSS080310:112544+182317 & 11:25:44.03 & +18:23:17.1 & 54535.36676 & 20.0 & $>$23 & \nodata & \nodata\nl  
CSS080308:090908+111115 & 09:09:07.99 & +11:11:15.3 & 54533.20757 & 18.4 & 21.2 & \nodata & \nodata\nl
CSS080303:081413+271939 & 08:14:13.15 & +27:19:39.0 & 54528.11129 & 18.1 & 23\tablenotemark{$\star$} & \nodata & \nodata\nl
CSS080303:075520+203908 & 07:55:19.52 & +20:39:08.4 & 54528.10484 & 19.3 & $>$23 & 2008bb & II\nl
CSS080302:124132+332203 & 12:41:31.98 & +33:22:03.2 & 54527.29264 & 19.2 & 22.5\tablenotemark{$\star$} & \nodata & \nodata\nl
CSS080302:145726+232348 & 14:57:26.21 & +23:23:48.2 & 54527.47633 & 18.0 & $>$22 & 2008av & \nodata\nl
CSS080228:142427-063345 & 14:24:26.63 & -06:33:44.8 & 54524.48530 & 18.3 & 21\tablenotemark{$\star$} & \nodata & \nodata\nl
CSS080228:105301-075656 & 10:53:00.81 & -07:56:55.9 & 54524.30812 & 19.2 & $>$22 & \nodata & \nodata\nl
CSS080227:160549+192717 & 16:05:49.17 & +19:27:17.2 & 54523.47442 & 17.8 & 22 & 2008ba & Ia\nl
CSS080227:113034+130905 & 11:30:33.96 & +13:09:05.3 & 54523.31151 & 18.7 & $>$23 & 2008au & \nodata\nl
CSS080219:151457+234110 & 15:14:56.75 & +23:41:10.1 & 54515.45977 & 16.6 & $>$23 & \nodata & \nodata\nl  
CSS080218:134903+315237 & 13:49:03.35 & +31:52:37.1 & 54514.41836 & 19.0 & 21.8 & \nodata & \nodata\nl
CSS080213:024607-073834 & 02:46:07.23 & -07:38:33.8 & 54509.09948 & 17.3 & $>$22 & 2008al & II\nl
CSS080210:080306+115121 & 08:03:06.09 & +11:51:20.8 & 54506.19556 & 18.8 & $>$23 & \nodata & \nodata\nl
CSS080206:130745+060805 & 13:07:44.64 & +06:08:05.0 & 54502.43974 & 19.8 & $>$23 & \nodata & \nodata\nl 
CSS080111:070447+412518 & 07:04:47.34 & +41:25:18.0 & 54476.25764 & 19.0 & $>$21 & \nodata & \nodata\nl 
CSS080111:145436-051254 & 14:54:35.97 & -05:12:53.8 & 54476.49167 & 18.1 & 21\tablenotemark{$\star$} & \nodata & \nodata\nl  
CSS080101:035151-062700 & 03:51:50.64 & -06:27:00.5 & 54466.25782 & 18.3 & $>$23 & \nodata & \nodata\nl
CSS080101:085249-052247 & 08:52:48.90 & -05:22:46.8 & 54466.37420 & 19.1 & 20.5 & \nodata & \nodata\nl
CSS071230:082550+220041 & 08:25:50.48 & +22:00:41.3 & 54464.35765 & 17.6 & 21.3 & \nodata & \nodata\nl 
CSS071230:081807+200719 & 08:18:07.05 & +20:07:19.2 & 54464.32780 & 18.1 & 22 & \nodata & \nodata\nl
CSS071219:223508-011160 & 22:35:08.45 & -01:11:59.6 & 54453.08583 & 18.3 & $>$23 & 2007qv & II\nl
CSS071218:120153-185822 & 12:01:52.80 & -18:58:21.7 & 54452.53386 & 12.9 & 9.8 & 2007sr & Ia\nl
CSS071216:121630+102303 & 12:16:29.63 & +10:23:02.9 & 54450.44569 & 18.9 & 21.5 & \nodata & \nodata\nl
CSS071206:102130+200007 & 10:21:30.15 & +20:00:07.4 & 54440.42300 & 18.6 & 21\tablenotemark{$\star$} & \nodata & \nodata\nl
CSS071204:035956+233018 & 03:59:56.34 & +23:30:18.2 & 54438.21532 & 18.5 & $>$22 & \nodata & \nodata\nl
CSS071204:002043+102044 & 00:20:42.71 & +10:20:44.1 & 54438.13146 & 18.2 & $>$22 & \nodata & \nodata\nl 
CSS071117:224558-003854 & 22:45:58.17 & -00:38:54.3 & 54421.10263 & 18.6 & 20.7 & 2007pu & Ia\nl
CSS071112:012312+120818 & 01:23:12.27 & +12:08:18.0 & 54416.17262 & 19.0 & 22\tablenotemark{$\star$} & \nodata & \nodata\nl
CSS071111:095436+045612 & 09:54:35.98 & +04:56:11.8 & 54415.46121 & 19.0 & 22.6 & \nodata & \nodata\nl
CSS071109:003334+214127 & 00:33:34.36 & +21:41:26.7 & 54413.27788 & 18.8 & 16.1 & \nodata & \nodata\nl 
CSS071107:093912+301341 & 09:39:11.62 & +30:13:41.5 & 54411.43569 & 18.2 & 20.0 & \nodata & \nodata\nl 
CSS071103:224527+103933 & 22:45:27.41 & +10:39:32.7 & 54407.18815 & 17.8 & $>$21 & 2007nm & Ia\nl
CSS071102:024106-033814 & 02:41:05.71 & -03:38:13.6 & 54406.26445 & 19.6 & $>$22 & 2007no & Ia\nl 
CSS071102:024457-044918 & 02:44:57.22 & -04:49:17.7 & 54406.26445 & 18.8 & $>$22 & \nodata & \nodata\nl  
CSS071101:015607+203404 & 01:56:06.60 & +20:34:03.5 & 54405.26126 & 18.8 & $>$22 & \nodata & \nodata\nl
CSS070320:124616+112555 & 12:46:15.81 & +11:25:55.4 & 54179.37725 & 17.6 & 20.8 & 2006tf & IIn\tablenotemark{a}\nl
\enddata
\tablenotetext{$\dagger$}{Sand et al. 2008, http://www.cfa.harvard.edu/iau/cbet/001400/CBET001410.txt.
However, first announced and spectroscopically confirmed by the Supernova Factory as SNF20080510-001.}
\tablenotetext{a}{Found in archival data.}
\tablenotetext{$\star$}{Estimated r-band magnitude.}
\tablecomments{
Col. (1), CSS ID.
Cols. (2) \& (3), Right Ascension and Declination J2000.
Col. (4) Modified Julian Date of Detection.
Col. (5) Detection Magnitude.
Col. (6) Approximate source galaxy Gunn $r$ magnitude.
Col. (7) IAU Designation.
}
\end{deluxetable}

\LongTables
\begin{deluxetable}{lccccccc}
\tablecaption{Cataclysmic Variable Candidates.\label{tab2}}
\footnotesize
\tablewidth{0pt}
\tablehead{\colhead{ID} & \colhead{RA} & \colhead{Dec}  & \colhead{Date} & \colhead{Mag} & \colhead{$Mag_Q$} &
  \colhead{$FUV$} & \colhead{$NUV$}\\
& h.m.s & $\arcdeg$ $\arcmin$ $\arcsec$  & $MJD$ & $$ & $$ & & $$}
\startdata
CSS080514:162606+225044 & 16:26:05.71 & +22:50:44.4 & 54600.36523 & 18.5 & 22.5  & \nodata & \nodata\\  
CSS080514:151021+182303 & 15:10:20.74 & +18:23:02.5 & 54600.22210 & 17.8 & 20.7 & \nodata & \nodata\\  
CSS080513:164002+073208 & 16:40:02.11 & +07:32:07.8 & 54599.31544 & 18.0 & 21\tablenotemark{$\star$}   & 21.4 & \nodata\\  
CSS080512:064608+403305 & 06:46:08.23 & +40:33:05.1 & 54598.14377 & 16.2 & 21\tablenotemark{$\star$}   & \nodata & 22.7\\  
CSS080512:173860+344023 & 17:38:59.82 & +34:40:23.1 & 54598.35924 & 18.7 & $>$21 & \nodata & \nodata\\
CSS080511:212555-032406 & 21:25:55.07 & -03:24:05.9 & 54597.43009 & 17.8 & $>$21 & \nodata & 23.3\\
CSS080506:085409+201339 & 08:54:09.41 & +20:13:39.2 & 54592.15678 & 17.7 & 20.5 & 21.5 & 20.3\\  
CSS080505:105836+120049 & 10:58:35.93 & +12:00:48.5 & 54591.20280 & 18.4 & $>$23 & \nodata & \nodata\tablenotemark{a}\\  
CSS080505:163121+103134 & 16:31:20.89 & +10:31:33.9 & 54591.37168 & 14.2 & 19.0 & \nodata & \nodata \\ 
CSS080505:214804+080951 & 21:48:04.40 & +08:09:50.5 & 54591.43395 & 18.3 & 20.5 & 22.1  & 21.6 \\
CSS080502:141002-124809 & 14:10:02.21 & -12:48:08.7 & 54588.32033 & 16.2 & 21\tablenotemark{$\star$}   & \nodata & \nodata \\  
CSS080501:223058+210147 & 22:30:58.32 & +21:01:47.0 & 54587.46701 & 16.7 & 21\tablenotemark{$\star$}   &  20.9 & 21.4 \\  
CSS080428:162502+390926 & 16:25:01.72 & +39:09:26.3 & 54584.28928 & 13.5 & 17.1 & 18.2 & 17.9\tablenotemark{b}\\  
CSS080428:163805+083758 & 16:38:05.39 & +08:37:58.5 & 54584.43007 & 15.4 & 19.0 & 21.2 & 20.8\tablenotemark{c}\\  
CSS080428:160524+060816 & 16:05:24.14 & +06:08:15.8 & 54584.39153 & 19.4 & 22.5 & \nodata & \nodata \\  
CSS080427:153634+332852 & 15:36:34.41 & +33:28:52.1 & 54583.41750 & 15.2 & 19.0 & \nodata & \nodata\tablenotemark{d}\\  
CSS080427:124418+300401 & 12:44:17.88 & +30:04:01.2 & 54583.32956 & 15.0 & 18.5 & 18.7 & 18.2\\  
CSS080427:131626-151313 & 13:16:25.68 & -15:13:13.5 & 54583.23364 & 16.9 & 20\tablenotemark{$\star$} & 20.6 & 20.6 \\  
CSS080426:162657-002549 & 16:26:56.80 & -00:25:48.6 & 54582.41287 & 18.0 & 22.2 & \nodata & \nodata \\
CSS080425:141712-180328 & 14:17:11.98 & -18:03:27.7 & 54581.33298 & 15.0 & 20.5  & \nodata & \nodata\\  
CSS080425:143143-032520 & 14:31:43.05 & -03:25:20.5 & 54581.41153 & 18.5 & $>$21 & \nodata & \nodata\\  
CSS080424:160232+161732 & 16:02:32.15 & +16:17:32.5 & 54580.44400 & 18.4 & 21.8 & \nodata & \nodata\\  
CSS080424:155326+114437 & 15:53:25.67 & +11:44:36.8 & 54580.44238 & 18.2 & 23.4 & \nodata & \nodata\\  
CSS080417:081543-004209 & 08:15:43.16 & -00:42:08.6 & 54573.11929 & 17.8 & 21.8 & \nodata & \nodata\\  
CSS080416:080854+355053 & 08:08:53.73 & +35:50:53.3 & 54572.15297 & 16.4 & 19.6 & 19.8 & 20.0\\  
CSS080415:162012+115257 & 16:20:12.00 & +11:52:57.1 & 54571.41746 & 19.3 & 22.3 & \nodata & \nodata\\  
CSS080411:154258-020705 & 15:42:58.36 & -02:07:04.8 & 54567.39688 & 16.8 & 23 & \nodata & \nodata\\  
CSS080409:174714+150048 & 17:47:14.34 & +15:00:47.7 & 54565.37312 & 17.0 & 21\tablenotemark{$\star$} & \nodata & \nodata\\  
CSS080409:081419-005022 & 08:14:18.90 & -00:50:22.1 & 54565.14409 & 14.8 & 19.0 & 19.0 & 19.0\\  
CSS080406:075648+305805 & 07:56:48.02 & +30:58:05.4 & 54562.16871 & 17.0 & 21\tablenotemark{$\star$} & 21.3 & 21.6\\  
CSS080405:161851-052509 & 16:18:50.76 & -05:25:08.6 & 54561.43750 & 16.1 & 21\tablenotemark{$\star$} & \nodata & \nodata\\  
CSS080404:213309+155004 & 21:33:09.43 & +15:50:04.3 & 54560.50262 & 15.7 & $>$21 & 18.8 & 19.0 \\  
CSS080403:160004+331114 & 16:00:03.71 & +33:11:13.9 & 54559.36671 & 15.2 & 19.5 & 15.8 & 16.0\tablenotemark{e}\\  
CSS080401:085113+344449 & 08:51:13.43 & +34:44:48.7 & 54557.18634 & 16.4 & 20 & \nodata & \nodata\\  
CSS080401:153151+152447 & 15:31:50.80 & +15:24:46.8 & 54557.43700 & 18.9 & 22.6 & \nodata & \nodata\\
CSS080331:160205+031632 & 16:02:04.80 & +03:16:31.8 & 54556.32329 & 17.1 & 22.8 & \nodata & \nodata\\  
CSS080329:143500-004606 & 14:35:00.23 & -00:46:06.3 & 54554.41592 & 15.0 & 18.4 & 18.7 & 18.6\tablenotemark{f}\\  
CSS080324:122060-102735 & 12:20:59.77 & -10:27:35.1 & 54549.34105 & 16.8 & 19.7 & 18.3 & 18.6 \\
CSS080309:084358+425037 & 08:43:58.06 & +42:50:37.2 & 54534.25796 & 18.1 & 19.9 & 21.0 & 20.0\\  
CSS080309:070501+372505 & 07:05:01.06 & +37:25:05.2 & 54534.18927 & 18.1 & $>$22 & \nodata & 22.7\\  
CSS080307:090624-085141 & 09:06:23.52 & -08:51:40.5 & 54532.17312 & 16.5 & $>$21 & \nodata & \nodata\\  
CSS080306:082655-000733 & 08:26:54.68 & -00:07:32.9 & 54531.19113 & 16.4 & 19.5 & \nodata & \nodata\tablenotemark{g}\\  
CSS080305:102938+414046 & 10:29:37.71 & +41:40:46.4 & 54530.30353 & 17.3 & 22.3 & \nodata & \nodata\\  
CSS080305:080846+313106 & 08:08:46.19 & +31:31:06.1 & 54530.12195 & 14.8 & 18.7 & 19.4 & 19.9\tablenotemark{h}\\  
CSS080304:164002+073208 & 16:40:02.14 & +07:32:07.6 & 54529.44911 & 17.9 & $>$21 & 21.4 & \nodata\\  
CSS080304:090240+052501 & 09:02:39.70 & +05:25:00.6 & 54529.20407 & 16.2 & 23.1 & \nodata & 23.7\tablenotemark{i}\\
CSS080303:073921+222454 & 07:39:21.16 & +22:24:54.0 & 54528.10431 & 18.5 & 22.4 & \nodata & \nodata\\  
CSS080302:160845+220610 & 16:08:44.79 & +22:06:10.0 & 54527.50294 & 17.9 & 21.0 & \nodata & \nodata\\  
CSS080227:132103+015329 & 13:21:03.18 & +01:53:29.2 & 54523.43337 & 14.1 & 19.2 & \nodata & \nodata\tablenotemark{j}\\  
CSS080227:112634-100210 & 11:26:33.98 & -10:02:10.1 & 54523.35480 & 16.2 & 18.9 & \nodata & 20.2\tablenotemark{k}\\
CSS080211:024602+345508 & 02:46:02.38 & +34:55:08.2 & 54507.11196 & 15.0 & 19 & \nodata & \nodata\tablenotemark{l}\\  
CSS080209:090628+052657 & 09:06:28.25 & +05:26:56.9 & 54505.30194 & 15.6 & 18.5 & 16.4 & 16.2\tablenotemark{m}\\  
CSS080209:084555+033929 & 08:45:55.05 & +03:39:29.3 & 54505.26151 & 18.5 & 20.9 & \nodata & \nodata\tablenotemark{n}\\  
CSS080208:103317+072119 & 10:33:17.25 & +07:21:18.7 & 54504.37302 & 15.5 & 19.8 & 20.2 & 20.0\\  
CSS080207:060038-080945 & 06:00:37.59 & -08:09:44.7 & 54503.20131 & 16.3 & 19.8  & \nodata & \nodata \\  
CSS080207:052034-000530 & 05:20:33.85 & -00:05:30.1 & 54503.20510 & 16.8 & 19.7 & \nodata & \nodata\\  
CSS080202:081415+080450 & 08:14:14.92 & +08:04:50.2 & 54498.23365 & 18.5 & 22 & \nodata & \nodata\\  
CSS080202:084608+011743 & 08:46:07.73 & +01:17:42.9 & 54498.23741 & 18.2 & 22.5 & 23.4 & \nodata\\  
CSS080201:115330+315836 & 11:53:30.23 & +31:58:36.0 & 54497.33687 & 17.5 & 19.9 & \nodata & \nodata\\  
CSS080131:163943+122414 & 16:39:42.70 & +12:24:14.4 & 54496.51670 & 17.4 & 19.2 & \nodata & \nodata\\  
CSS080130:033056+251255 & 03:30:55.66 & +25:12:55.2 & 54495.15831 & 16.5 & 21\tablenotemark{$\star$} & \nodata & \nodata\\  
CSS080130:090951+184947 & 09:09:50.52 & +18:49:47.1 & 54495.31194 & 12.6 & 15.7 & 17.0 & 17.0\tablenotemark{o}\\ 
CSS080130:021110+171624 & 02:11:10.22 & +17:16:24.3 & 54495.13080 & 14.3 & 19.0 & 19.8 & 19.8 \\  
CSS080130:105550+095621 & 10:55:50.08 & +09:56:20.5 & 54495.44905 & 15.4 & 18.5 & \nodata & \nodata\\
CSS080110:032627+070744 & 03:26:27.26 & +07:07:44.4 & 54475.19676 & 17.7 & 21 & \nodata & \nodata\\  
CSS080101:232519-081819 & 23:25:19.20 & -08:18:18.8 & 54466.08479 & 14.9 & 18.6 & 19.9 & 19.3\tablenotemark{p}\\  
CSS071231:082822+105344 & 08:28:21.75 & +10:53:44.5 & 54465.38067 & 17.4 & 22.2 & 22.0 & \nodata\\  
CSS071215:041456+215643 & 04:14:55.71 & +21:56:43.1 & 54449.23934 & 17.1 & 21\tablenotemark{$\star$} & \nodata & \nodata\\ 
CSS071214:090904+091714 & 09:09:04.38 & +09:17:13.5 & 54448.44999 & 16.9 & 22 & \nodata & \nodata\\  
CSS071116:214843-000723 & 21:48:42.53 & -00:07:23.5 & 54420.09361 & 15.8 & 22.4 & 23.0 & 23.3\\  
CSS071115:172406+411410 & 17:24:06.32 & +41:14:10.1 & 54419.06981 & 17.1 & 18.5 & 19.7 & 19.1\tablenotemark{q}\\ 
CSS071115:044216-002334 & 04:42:16.03 & -00:23:33.9 & 54419.28884 & 17.0 & 22\tablenotemark{$\star$} & \nodata & 21.9\\  
CSS071112:024457+352249 & 02:44:57.42 & +35:22:49.3 & 54416.20408 & 18.4 & 21\tablenotemark{$\star$} & \nodata & \nodata\\  
CSS071112:085823-003729 & 08:58:22.86 & -00:37:29.0 & 54416.42038 & 17.5 & 22.0 & \nodata & \nodata\\ 
CSS071106:075747+305307 & 07:57:46.97 & +30:53:07.2 & 54410.40003 & 18.5 & $>$23 & \nodata & \nodata\\  
\enddata

\tablenotetext{$\star$}{Estimated r-band magnitude.}
\tablenotetext{a}{Mahabal et al.~(2008), Atel 1520}
\tablenotetext{b}{V844 Her}
\tablenotetext{c}{V544 Her}
\tablenotetext{d}{SDSS J153634+332851}
\tablenotetext{e}{VW CrB}
\tablenotetext{f}{OU Vir}
\tablenotetext{g}{SDSS QSO candidate J082654.69-000733.1 (Richards et al. 2004).}
\tablenotetext{h}{Cnc SDSS J080846+313106}
\tablenotetext{i}{Djorgovski et al.~(2008), Atel 1418 and 1411}
\tablenotetext{j}{HV-Vir}
\tablenotetext{k}{Eclipsing system}
\tablenotetext{l}{Antipin V80}
\tablenotetext{m}{Hya SDSS J090628+052656}
\tablenotetext{n}{First appeared in CBAT Jan 20th.}
\tablenotetext{o}{GY Cnc}
\tablenotetext{p}{EG Aqr}
\tablenotetext{q}{V1007 Her}


\tablecomments{
Col. (1), CSS ID.
Cols. (2) \& (3), Right Ascension and Declination J2000.
Col. (4) Modified Julian Date of detection.
Col. (5) Catalina detection magnitude.
Col. (6) Approximate Gunn r-band quiescent magnitude.
Col. (7) \& (8), GALEX, FUV and NUV magnitudes.
}
\end{deluxetable}

\begin{deluxetable}{lccccccccc}
\tablecaption{Blazar Candidates.\label{tab3}}
\footnotesize
\tablewidth{0pt}
\tablehead{\colhead{ID} & \colhead{RA} & \colhead{Dec}  & \colhead{Date} & \colhead{u} &  \colhead{g} & \colhead{r} & \colhead{i} & \colhead{z} & \colhead{CSS}\\
& h.m.s & $\arcdeg$ $\arcmin$ $\arcsec$  & $MJD$ & $$ & $$ & & $$}
\startdata
CSS080506:120952+181007 & 12:09:51.75 & +18:10:07.0 & 54592.21514 & 19.3 & 18.9 & 18.6 & 18.3 & 18.1 & 15.8\tablenotemark{a}\\  
CSS080409:154159+235603 & 15:41:59.97 & +23:56:03.3 & 54565.28634 & 23.9 & 23.7 &  23.6 & 22.6&  22.7 & 18.8\tablenotemark{a}\\
CSS080426:165347+164950 & 16:53:46.61 & +16:49:49.5 & 54582.44017 & 22.4 & 22.2 & 22.4 & 22.2 &  21.5 & 19.0\\
CSS080306:141549+090354 & 14:15:48.80 & +09:03:54.4 & 54531.32834 & 21.2 & 20.5 & 20.4 & 19.9 &  19.6 & 18.6\\
CSS080208:120722+250650 & 12:07:21.938 & +25:06:50.26 & 54504.32945 & \nodata & \nodata & \nodata & \nodata & \nodata & 19.3\\
\enddata
\tablenotetext{a}{Spectroscopic candidates (Mahabal et al.~2008).}

\tablecomments{
Col. (1), CSS ID.
Cols. (2) \& (3), Right Ascension and Declination J2000.
Col. (4) Modified Julian Date of detection.
Col. (5 - 9), SDSS DR6 magnitudes.
Col. (10) Catalina detection magnitude.
}
\end{deluxetable}

\end{document}